\documentstyle[12pt,epsf,epsfig,fleqn]{article}

\newcommand{\be}{\begin{equation}}
\newcommand{\ee}{\end{equation}}
\newcommand{\lee}[1]{\label{#1} \end{equation}}
\newcommand{\bea}{\begin{eqnarray}}
\newcommand{\leea}[1]{\label{#1} \end{eqnarray}}
\newcommand{\eea}{\end{eqnarray}}
\newcommand{\nn}{\nonumber}
\newcommand{\eq}[1]{eq.~(\ref{#1})}

\newcommand{\eqs}[2]{eqs.~(\ref{#1}) and (\ref{#2})}

\newcommand{\fig}[1]{fig.~(\ref{#1})}
\newcommand{\loadeps}[1]{\epsfig{file=#1.eps,width=45mm}}

\newcommand{\ga}{\gamma}
\newcommand{\de}{\delta}

\newcommand{\la}{\lambda}

\newcommand{\De}{\Delta}
\newcommand{\La}{\Lambda}

\begin{document}
\title{Flow equations in the light-front
perturbation theory
\vspace*{.5cm}}
\author{E.~L.~Gubankova\thanks{E-mail address: 
elena@frodo.tphys.uni-heidelberg.de}
\vspace*{.5cm}\\
\normalsize\it Institut f\"ur Theoretische Physik der Universit\"at 
Heidelberg \\
\normalsize\it Philosophenweg 19, D69120 Heidelberg, FRG}
\date{}

\maketitle

\vspace*{1cm}
\begin{abstract}

The method of flow equations is applied to QED 
in the light-front dynamics. To second order in the coupling
the particle number conserving part of the effective 
QED Hamiltonian has two terms of different structure.
The first term gives the Coulomb interaction
and the correct spin splittings of positronium;
the contribution of the second term to mass spectrum
depends on the explicit form of unitary transformation
and may influence the spin-orbit.

\end{abstract}


\newpage
\section{Introduction}

In the previous work \cite{GuWe} we have outlined the strategy
to construct an effective renormalized Hamiltonian 
by means of flow equations. Requiring that the particle number
(Fock state) conserving terms in the Hamiltonian were considered
to be diagonal and the other terms off-diagonal
the block-diagonal effective Hamiltonian was obtained.
The main advantage of this procedure is, that finally states
of different particle number (Fock sectors) are completely decoupled
in the effective Hamiltonian, and thus the bound state problem,
which is in general a many-body one, reduces to a few-body problem.
For the positronium this means, that one is left with a two-particle
problem, since the particle number violating contributions
are eliminated.

We perform the unitary transformation to bring
the bare cutoff Hamiltonian of the gauge field theory
to a block-diagonal form. We distinguish therefore
the 'diagonal'-particle number (Fock state) conserving
and the 'rest'-particle number (Fock state) changing sectors.
The elimination of the "rest" part of Hamiltonian 
generates the new interactions in the "diagonal" sectors.

We study QED on the light-front in this approach.
We use one of the most convenient prescription
of light-front perturbation theory as formulated
by Brodsky and Lepage \cite{LeBr}. In second order in coupling
there is a new interaction between electron and positron
that arise from the eliminating matrix elements
of electron-photon vertex. 

Making use of Brodsky, Lepage light-front perturbation theory
it is possible to separate two different structures
in the effective electron-positron interaction.
The contribution of both of them to the mass spectrum is considered.

\section{Flow equations and the effective Hamiltonian}
\label{Flow equations}

Flow equations perform the unitary transformation, which brings
the Hamiltonian to a block-diagonal form with the number of particles
(or Fock state) conserving in each block. In what follows we
distinguish between the 'diagonal' (here Fock state conserving)
and 'rest' (Fock state changing) sectors of the Hamiltonian.
We break the Hamiltonian as
\bea
&& H=H_{0d}+H_d+H_r
\leea{f1}
where $H_{0d}$ is the free Hamiltonian; and the indices 'd','r'
correspond to 'diagonal','rest' parts of the Hamiltonian, respectively.
The flow equation for the Hamiltonian and the generator 
of unitary transformation are written \cite{We}
\bea
&& \frac{dH}{dl}=[\eta,H_d+H_r]+[[H_d,H_r],H_{0d}]
+[[H_{0d},H_r],H_{0d}]\nn\\
&& \eta=[H_{0d},H_r]+[H_d,H_r]
\leea{f2}
In the basis of the eigenfunctions of the free Hamiltonian 
\bea
&& H_{0d}|i>=E_i|i>
\leea{f3}
one obtains for the matrix-elements between the many-particle states
\bea
&& \frac{dH_{ij}}{dl}=[\eta,H_d+H_r]_{ij}-(E_i-E_j)[H_d,H_r]_{ij}
-(E_i-E_j)^2H_{rij}\nn\\
&& \eta_{ij}=(E_i-E_j)H_{rij}+[H_d,H_r]_{ij}
\leea{f4}
The energy differences are given by
\bea
&& E_i-E_j=\sum_{k=1}^{n_2}E_{i,k}-\sum_{k=1}^{n_1}E_{j,k}
\leea{f5}
where $E_{i,k}$ and $E_{j,k}$ are the energies of the created and
annihilated particles, respectively. 
The generator belongs to the 'rest' sector, i.e.
$\eta_{ij}=\eta_{rij}, \eta_{dij}=0$.
In what follows we use
\bea
&& [\hat{O}_r,\hat{H}_d]_d=0\nn\\
&& [\hat{O}_r,\hat{H}_d]_r\neq 0
\leea{}
where $\hat{O}_r$ is the operator from the 'rest' sector
(for example $\hat{H}_r$ or $\hat{\eta}_r$)
and $\hat{H}_d$ is the diagonal part of Hamiltonian.

For the 'diagonal' $(n_1=n_2)$
and 'rest' $(n_1\neq n_2)$
sectors of the Hamiltonian one has correspondingly
\bea
&& \frac{dH_{dij}}{dl}=[\eta,H_r]_{dij} \nn\\
&& \frac{dH_{rij}}{dl}=[\eta,H_d+H_r]_{rij}-(E_i-E_j)[H_d,H_r]_{rij}
+\frac{du_{ij}}{dl}\frac{H_{rij}}{u_{ij}}
\leea{f6}
where we have introduced the cutoff function $u_{ij}(l)$
\bea
&& u_{ij}(l)={\rm e}^{-(E_i-E_j)^2l}
\leea{f7}
For simplicity we neglect the dependence of the energy $E_i$
on the flow parameter $l$.
The main difference between these two sectors is the presence
of the third term in the 'rest' sector 
$\frac{du_{ij}}{dl}\frac{H_{rij}}{u_{ij}}$,
which insures the band-diagonal structure for the 'rest' part
\bea
&& H_{rij}=u_{ij}\tilde{H}_{rij}
\leea{f8}
i.e. in the 'rest' sector only the matrix elements with the
energy difference $|E_i-E_j|<1/\sqrt{l}$ are not zero.
In the similarity renormalization scheme \cite{GlWi}
the width of the band corresponds to the UV cutoff $\la$.
The connection between the two quantities is given
\bea
&& l=\frac{1}{\la^2}
\leea{f9}
The matrix elements of the interactions, which change
the Fock state, are strongly suppressed,if the energy difference
exceeds $\la$, while for the Fock state conserving part
of the Hamiltonian the matrix elements with all energy differences
are present.
As the flow parameter $l\rightarrow\infty$ (or $\la\rightarrow 0$)
the 'rest' part is completely eliminated, except maybe
for the matrix elements with $i=j$. One is left with the block-diagonal
effective Hamiltonian.

Generally, the flow equations are written
\bea
&& \frac{dH_{ij}}{dl}=[\eta,H_d+H_r]_{ij}-(E_i-E_j)[H_d,H_r]_{ij}
+\frac{du_{ij}}{dl}\frac{H_{ij}}{u_{ij}}\nn\\
&& \eta_{ij}=[H_d,H_r]_{ij}
+\frac{1}{E_i-E_j}\left(-\frac{du_{ij}}{dl}\frac{H_{ij}}{u_{ij}}\right)
\leea{f10}
where the following condition on the cutoff function
in 'diagonal' and 'rest' sectors, respectively, is imposed
\bea
&& u_{dij}=1\nn\\
&& u_{rij}=u_{ij}
\leea{f11}
One recovers with this condition the flow equations \eq{f6}
for both sectors.
Other unitary transformations, which bring the Hamiltonian 
to the block-diagonal form, with the Fock state conserving in each block,
are used \cite{GlWi}                        
\bea
&& \frac{dH_{ij}}{d\la}=u_{ij}[\eta,H_d+H_r]_{ij}
+r_{ij}\frac{du_{ij}}{d\la}\frac{H_{ij}}{u_{ij}}\nn\\
&& \eta_{ij}=\frac{r_{ij}}{E_i-E_j}
\left([\eta,H_d+H_r]_{ij}-\frac{du_{ij}}{d\la}\frac{H_{ij}}{u_{ij}}\right)
\leea{f12}
and \cite{WiPe}
\bea
&& \frac{dH_{ij}}{d\la}=u_{ij}[\eta,H_d+H_r]_{ij}
+\frac{du_{ij}}{d\la}\frac{H_{ij}}{u_{ij}}\nn\\
&& \eta_{ij}=\frac{1}{E_i-E_j}
\left(r_{ij}[\eta,H_d+H_r]_{ij}
-\frac{du_{ij}}{d\la}\frac{H_{ij}}{u_{ij}}\right)
\leea{f13}
where $u_{ij}+r_{ij}=1$; and the constrain \eq{f11}
on the cutoff function in both sectors is implied.
One can choose the sharp cutoff function
$u_{ij}=\theta(\la-|\De_{ij}|)$.

We consider the flow equations in the perturbative frame,
therefore we break the Hamiltonian
\bea
&& H=H_{0d}+\sum_n(H_{d}^{(n)}+H_{r}^{(n)})
\leea{f15}
where $H^{(n)}\sim e^n$, $e$ is the bare coupling constant
(here we do not refer to the definite field theory).
To the order of $n$ in coupling constant
the flow equations in both sectors are given
\bea
&& \frac{dH_{dij}^{(n)}}{dl}=\sum_k[\eta^{(k)},H_r^{(n-k)}]_{dij}\nn\\
&& \eta_{dij}^{(n)}=0 \nn\\
&& \frac{dH_{rij}^{(n)}}{dl}=\sum_k[\eta^{(k)},H_d^{(n-k)}
+H_r^{(n-k)}]_{rij}
-(E_i-E_j)\sum_k[H_d^{(k)},H_r^{(n-k)}]_{rij}
+\frac{du_{ij}}{dl}\frac{H_{rij}^{(n)}}{u_{ij}}\nn\\
&& \eta_{rij}^{(n)}=\sum_k[H_d^{(k)},H_r^{(n-k)}]_{rij}
+\frac{1}{E_i-E_j}\left(-\frac{du_{ij}}{dl}\frac{H_{rij}^{(n)}}
{u_{ij}}\right)\nn\\
&& \eta^{(n)}=\eta^{(n)}_d+\eta^{(n)}_r
\leea{f16}
We solve these equations in the leading order for the Fock state
conserving sector.
\bea
&& \frac{dH_{dij}^{(2)}}{dl}=[\eta_r^{(1)},H_r^{(1)}]_{dij}\nn\\
&& \eta_{rij}^{(1)}=-\frac{1}{E_i-E_j}\frac{dH_{rij}^{(1)}}{dl}\nn\\
&& \frac{dH_{rij}^{(1)}}{dl}=\frac{du_{ij}}{dl}\frac{H_{rij}^{(1)}}{u_{ij}}
\leea{f17}
and $H_{rij}^{(1)}=H_{rji}^{(1)}$.
Explicitly one has
\bea
&& \frac{dH_{dij}^{(2)}}{dl}=-\sum_k\left(
\frac{1}{E_i-E_k}\frac{dH_{rik}^{(1)}}{dl}H_{rjk}^{(1)}
+\frac{1}{E_j-E_k}\frac{dH_{rjk}^{(1)}}{dl}H_{rik}^{(1)}
\right)_d\nn\\
&& H_{rij}^{(1)}(l)=H_{rij}^{(1)}(l=0)\frac{f_{ij}(l)}{f_{ij}(l=0)}
\leea{f18}
where we have introduced the function $f_{ij}$ defining the leading order
solution for the 'rest' part. Further we refer to it as similarity function.
Here
\bea
&& f_{ij}(l)=u_{ij}(l)={\rm e}^{-(E_i-E_j)^2l}
\leea{f19}
The similarity function $f_{\la}(\De)$ has the same behavior
(when $\la\rightarrow\infty$~ $f_{\la}(\De)=1$,
and when $\la\rightarrow 0$~ $f_{\la}(\De)=0$)
as the cutoff function $u_{\la}(\De)$; in the leading order
the similarity function shows how fast the 'rest' sector
is eliminated.

Making use of the connection $l=1/\la^2$, we get
\bea
&& \frac{dH_{dij}^{(2)}}{d\la}=-\sum_k\left(
\frac{1}{E_i-E_k}\frac{dH_{rik}^{(1)}}{d\la}H_{rjk}^{(1)}
+\frac{1}{E_j-E_k}\frac{dH_{rjk}^{(1)}}{d\la}H_{rik}^{(1)}
\right)_d\nn\\
&& H_{rij}^{(1)}(\la)=H_{rij}^{(1)}(\La\rightarrow\infty)
\frac{f_{ij}(\la)}{f_{ij}(\La\rightarrow\infty)}
\leea{f20}
The equations \eq{f20} are the same also for the unitary transformations 
given above \eqs{f12}{f13} up to the choice of the function $f_{ij}(\la)$.
Neglecting the dependence of the energy $E_i$ on the cutoff, 
one has
\bea
&& H_{dij}^{(2)}(\la)=H_{dij}^{(2)}(\La\rightarrow\infty)
+\left(\sum_k H_{rik}^{(1)}(\La\rightarrow\infty)
H_{rjk}^{(1)}(\La\rightarrow\infty)\right)_d\nn\\
&& \times\left(
\frac{1}{E_i-E_k}\int_{\la}^{\infty}
\frac{df_{ik}(\la')}{d\la'}f_{jk}(\la')d\la'
+\frac{1}{E_j-E_k}\int_{\la}^{\infty}
\frac{df_{jk}(\la')}{d\la'}f_{ik}(\la')d\la'
\right)_d
\leea{f21}
where $\La$ is the bare cutoff; the sum $\sum_k$ is over
all intermediate states; and the label 'd' denotes the 'diagonal' sector.
Specifying the function $f_{ij}$ we get the interaction
$H_d^{(2)}$, generated by different unitary transformations.
One has corresponding\\
for the flow equations \eq{f10}
\bea
&& f_{ij}(\la)=u_{ij}(\la)={\rm e}^{-\frac{\De_{ij}^2}{\la^2}}
\leea{f22}
and for the unitary transformations \eqs{f12}{f13} 
\bea
&& f_{ij}(\la)=u_{ij}(\la){\rm e}^{r_{ij}(\la)}\nn\\
&& f_{ij}(\la)=u_{ij}(\la)\nn\\
&& u_{ij}=\theta(\la-|\De_{ij}|),~~u_{ij}+r_{ij}=1
\leea{f23}
where $\De_{ij}=\sum_{k=1}^{n_2}E_{i,k}-\sum_{k=1}^{n_1}E_{j,k}$.

To illustrate the method we consider QED on the light-front 
in the next section. We calculate the generated interaction \eq{f21}
in the electron-positron sector to solve for 
the positronium mass spectrum.
In this case $H_r^{(1)}(\La\rightarrow\infty)$ is 
the electron-photon vertex with the bare coupling constant $e$;
and the initial value of generated interaction is 
$H_d^{(2)}(\La\rightarrow\infty)=0$.

\section{Renormalized effective electron-positron interaction}
\label{Renormalized interaction}

In this section we give the effective Hamiltonian in the light front
dynamics for the positronium system, generated by the unitary transformation
\cite{GuWe}.

The light front Schr\"odinger equation for the positronium model reads
\bea
&& H_{LC}|\psi_n>=M_n^2|\psi_n>
\leea{r1}
where $H_{LC}=P^{\mu}P_{\mu}$ is the invariant mass (squared) operator, 
referred for convenience as the light front Hamiltonian of positronium 
and $|\psi_n>$ is the corresponding eigenfunction. 

The canonical Hamiltonian of the system
$H_{LC}$ is in general an infinite dimensional matrix, namely
contains infinite many Fock sectors (i.e. one has for the positronium
wave function $|\psi_n>=c_{e\bar{e}}|(e\bar{e})_n>
+c_{e\bar{e}\ga}|(e\bar{e}\ga)_n>
+c_{e\bar{e}\ga\ga}|(e\bar{e}\ga\ga)_n>+...$) and the states 
with infinite large energies. We obtain a finite dimensional Hamiltonian by:

$(1)$ introducing the bare cutoff (regularization) with the result
$H_{LC}^B(\La)$ - the bare Hamiltonian;

$(2)$ performing the unitary transformation 
with the result $H_{LC}^{eff}$ - the effective renormalized
Hamiltonian;

$(3)$ truncating the Fock space to the lowest Fock sector ($|e\bar{e}>$)
with the result $\tilde{H}_{LC}^{eff}$ - the effective 
renormalized Hamiltonian acting in the electron-positron sector.

\begin{figure}
\setlength{\unitlength}{1mm}
\begin{picture}(170,71)
\put(19,36){\makebox(56,34.61){ \loadeps{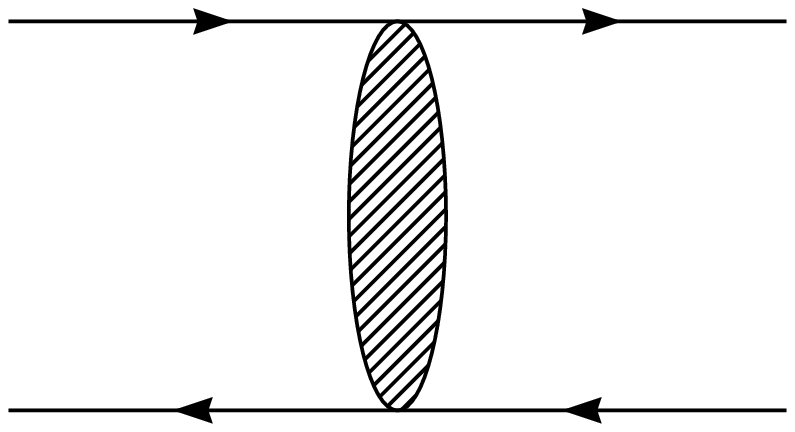} }}
\put(18,67){$p'_1\;(x',\vec{k'}_{\perp})$}
\put(50,67){$p_1\;(x,\vec{k}_{\perp})$}
\put(18,38){$p'_2\;(1\!-\!x',-\vec{k'}_{\perp})$}
\put(50,38){$p_2\;(1\!-\!x,-\vec{k}_{\perp})$}
  \put(75,36){\makebox(56,34.61){ \loadeps{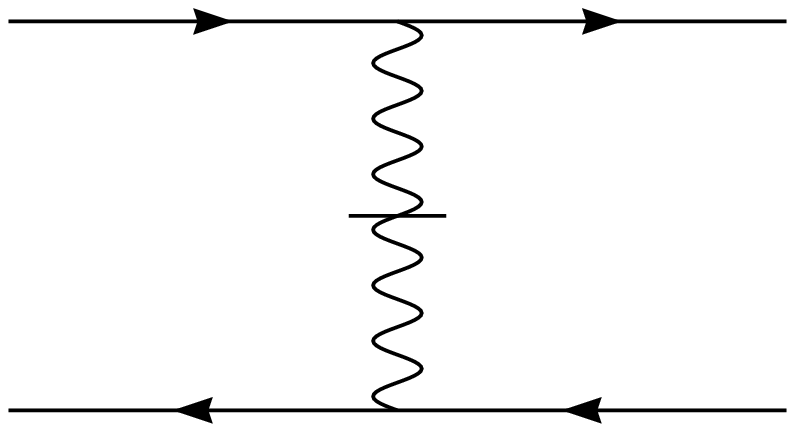} }}
\put(73,53.305){$+$}
\put(38,0){\makebox(56,34.61){ \loadeps{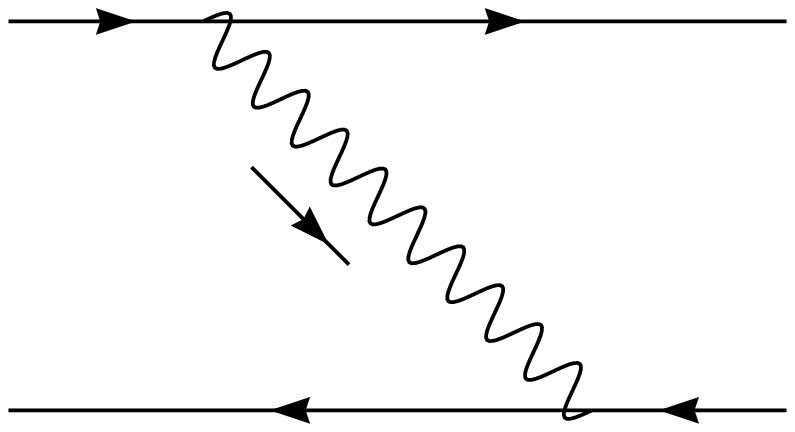} }}
  \put(94,0){\makebox(56,34.61){ \loadeps{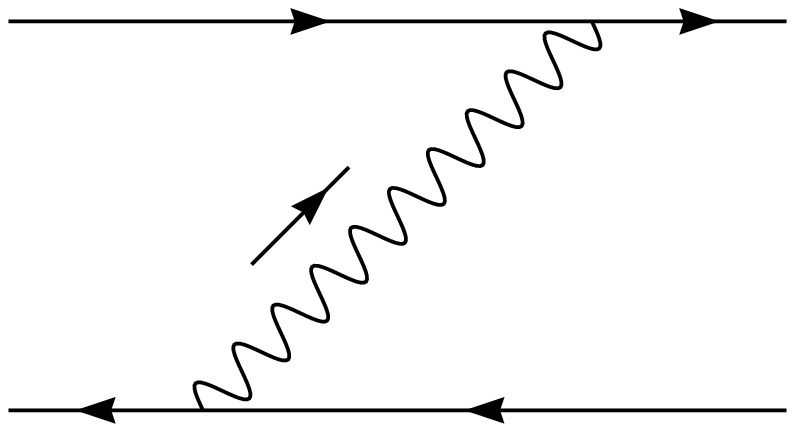} }}
\put(36,17.305){$+$}
\put(92,17.305){$+$}
\end{picture}
\vspace{3cm}
\caption{The effective electron-positron interaction 
in the exchange channel; 
the diagrams correspond to the generated and instantaneous interactions.
The perturbative photon exchange with two different time orderings
is also depicted.}
\label{1}
\end{figure}

The eigenvalue equation is written then
\bea
&& \tilde{H}_{LC}^{eff}|(e\bar{e})_n>=M_n^2|(e\bar{e})_n>
\leea{r2}
where $n$ labels all quantum numbers.
The effective light front Hamiltonian is splitted into
the free (noninteracting) part and effective electron-positron interaction
\bea
&& \tilde{H}_{LC}^{eff}=H_{LC}^{(0)}+V_{LC}^{eff}
\leea{r3}
The continuum version of the light front equation \eq{r2} 
is then expressed by the integral equation (Jacobi momenta are introduced
on \fig{1})
\bea
&& \hspace{-2cm} \left(M_n^2-\frac{m^2+\vec{k}_{\perp}^{2}}{x(1-x)}\right)
\psi_n(x,\vec{k}_{\perp};\la_1,\la_2)\nn\\
&& \hspace{1.5cm} =\sum_{\la^{'}_1,\la^{'}_2}\int_D 
\frac{dx' d^2 \vec{k}_{\perp}^{'}}{2(2\pi)^3}
<x,\vec{k}_{\perp};\la_1,\la_2|
V_{LC}^{eff}|x',\vec{k}_{\perp}^{'};\la_1^{'},\la_2^{'}>
\psi_n(x',\vec{k}_{\perp}^{'};\la_1^{'},\la_2^{'})
\leea{r4}
where the wave function is normalized
\bea
&& \sum_{\la_1,\la_2} 
\frac{dx d^2 \vec{k}_{\perp}}{2(2\pi)^3}
\psi_n^{*}(x,\vec{k}_{\perp};\la_1,\la_2)
\psi_{n'}(x,\vec{k}_{\perp};\la_1,\la_2)=\de_{nn'}
\leea{r5}
The integration domain $D$ is restricted by the covariant cutoff
condition of Brodsky and Lepage \cite{LeBr}
\bea
&& \frac{m^2+\vec{k}_{\perp}^{2}}{x(1-x)}\leq \La^2+4m^2
\leea{r6}
which allows for states to have the kinetic energy below the bare 
cutoff $\La$.
 
For the effective electron-positron interaction one has 
in the exchange and annihilation channels 
\bea
&& V_{LC}^{eff}=V_{exch}+V_{ann}=
\sum_{channel}\lim_{\la\rightarrow 0 }
(V_{\la}^{gen}+V_{\la}^{PT}+V_{\la}^{inst})
\leea{r7}
where the terms in \eq{r7} correspond to the interaction
generated by the unitary transformation \eq{f21},
the perturbative photon exchange 
with the energy below the cutoff 
and the instantaneous exchange term arising in the light-front gauge,
respectively.
Here $\la$ is the 'running' cutoff, that defines the 'continuous' step
of the unitary transformation $U(\la,\La)$ and related to 
the flow parameter $l$ as $l=1/\la^2$.

Following the rules of light-front perturbative theory \cite{LeBr}
one has for the interaction \eq{f21}, generated in the second order
$O(e^2)$ in the electron-positron sector
\bea
&& \hspace{-1cm} 
V_{\la}^{gen}=-e^2<\ga^{\mu}\ga^{\nu}>\nn\\
&& \hspace{-1.5cm}
\times [\frac{\theta(q^+)}{q^+}D_{\mu\nu}(q)
(\frac{1}{D_1}\int_{\la}^{\infty}
\frac{df_{\la'}(D_1)}{d\la'}f_{\la'}(D_2)d\la'
+\frac{1}{D_2}\int_{\la}^{\infty}
\frac{df_{\la'}(D_2)}{d\la'}f_{\la'}(D_1)d\la')\nn\\
&& \hspace{-1.5cm}
+\frac{\theta(-q^+)}{-q^+}D_{\mu\nu}(-q)
(\frac{1}{-D_1}\int_{\la}^{\infty}
\frac{df_{\la'}(-D_1)}{d\la'}f_{\la'}(-D_2)d\la'
+\frac{1}{-D_2}\int_{\la}^{\infty}
\frac{df_{\la'}(-D_2)}{d\la'}f_{\la'}(-D_1)d\la')]\nn\\
&&
\leea{r8}
where we sum ($\sum_k$ in \eq{f21}) the two terms corresponding to the two
time-ordered diagrams;
$f_{\la}(\De)$ is the similarity function, arising 
from the unitary transformation and specified below; 
$D_{\mu\nu}(q)=\frac{q^{\perp 2}}{q^{+ 2}}\eta_{\mu}\eta_{\nu}
+\frac{1}{q^+}(\eta_{\mu}q^{\perp}_{\nu}+\eta_{\nu}q^{\perp}_{\mu})
-g^{\perp}_{\mu\nu}$ is the photon propagator in light-front gauge
\cite{BaNaSo}, $\eta_{\mu}=(0,\eta_{+}=1,0,0)$; $D_1,D_2$ and $D$ are 
energy denominators given below; $q$ is the exchanged momentum.
The notation $<\ga^{\mu}\ga^{\nu}>$ is introduced for the matrix element
given in exchange channel as \fig{1}
\bea
&& <\ga^{\mu}\ga^{\nu}>|_{exch}
=\frac{\bar{u}(p_1,\la_1)}{\sqrt{p_1^{+}}}\ga^{\mu}
\frac{u(p_1^{'},\la_1^{'})}{\sqrt{p_1^{' +}}}
\frac{\bar{v}(p_2^{'},\la_2^{'})}{\sqrt{p_2^{' +}}}\ga^{\nu}
\frac{v(p_2,\la_2)}{\sqrt{p_2^{+}}}
P^{+ 2}
\leea{r12}
where $p_i,p_i^{'}$ are light-front three-momenta carried by the constituents,
$\la_i,\la_i^{'}$ are their light-front helicities,
$u(p_1,\la_1),v(p_2,\la_2)$ are their spinors \cite{LeBr};
index $i=1,2$ refers to electron and positron, respectively;
$P=(P^+,P^{\perp})$ is light-front positronium momentum.
(For the annihilation channel see Appendix B).

Making use of the symmetry
\bea
&& f_{\la}(-D)=f_{\la}(D)\nn\\
&& D_{\mu\nu}(-q)=D_{\mu\nu}(q)
\leea{r9}
we have the following electron-positron interaction
generated by the unitary transformation
\bea
&& \hspace{-2cm} V_{\la}^{gen}=-e^2<\ga^{\mu}\ga^{\nu}>
\frac{1}{q^+}D_{\mu\nu}(q)
\left(
\frac{1}{D_1}\int_{\la}^{\infty}
\frac{df_{\la'}(D_1)}{d\la'}f_{\la'}(D_2)d\la'
+\frac{1}{D_2}\int_{\la}^{\infty}
\frac{df_{\la'}(D_2)}{d\la'}f_{\la'}(D_1)d\la'
\right)\nn\\
&&
\leea{r10}
We combine all the interactions \eq{r7} together \cite{GuWe}
\bea
&& \hspace{-2cm} V_{\la}^{gen} = -e^2<\ga^{\mu}\ga^{\nu}>
\frac{1}{q^+}D_{\mu\nu}(q)
\left(\frac{1}{D_1}
\int_{\la}^{\infty}\frac{df_{\la'}(D_1)}{d\la'}f_{\la'}(D_2)d\la'+
\frac{1}{D_2}
\int_{\la}^{\infty}\frac{df_{\la'}(D_2)}{d\la'}f_{\la'}(D_1)d\la'
\right)\nn\\
&& \hspace{-2cm} V_{\la}^{PT} = -e^2<\ga^{\mu}\ga^{\nu}>
\frac{1}{q^+}D_{\mu\nu}(q)
\frac{1}{D}f_{\la}(D_1)f_{\la}(D_2)\nn\\
&& \hspace{-2cm} V^{inst} = -e^2<\ga^{\mu}\ga^{\nu}>
\frac{1}{q^{+2}}\eta_{\mu}\eta_{\nu}
\leea{r11}
The energy denominators in the generated interaction $V_{\la}^{gen}$
and the exchanged momentum are\\
in the {\bf exchange channel}
\bea
&& D_1^{exch}=\De_{p'_1,p_1};~~
   D_2^{exch}=\De_{p_2,p'_2};~~
   q^{exch}=p'_1-p_1\equiv q
\leea{r13}
in the {\bf annihilation channel}
\bea
&& D_1^{ann}=\De_{p'_1,-p'_2};~~
   D_2^{ann}=\De_{p_2,-p_1};~~
   q^{ann}=p_1+p_2\equiv P
\leea{r14}
where the notation is introduced 
$\De_{p_1p_2}=p_1^--p_2^--(p_1-p_2)^-$. The energy denominator
in the perturbative term $V_{\la}^{PT}$ is given
\bea
&& D=\sum_{inc}p^- -\sum_{interm}p^-
\leea{r15}
where the sums are over the light-front energies, $p^-$,
of the incident (inc) and intermediate (interm) particles \cite{LeBr}.

In what follows we use Jacobi momenta \fig{1}
\bea
&& p_1(xP^+,x\vec{P}_{\perp}+\vec{k}_{\perp})\nn\\
&& p_2((1-x)P^+,(1-x)\vec{P}_{\perp}-\vec{k}_{\perp})
\leea{r16}
and corresponding for the momenta $p'_1,p'_2$;
here $x$ is the light-front fraction of electron momentum
and $P(P^+,\vec{P}_{\perp})$ is the total momentum of positronium.
For convenience introduce
\bea
&& D_1^{exch}=\frac{\De_1}{P^+}=-\frac{\tilde{\De}_1}{q^+};~~
D_2^{exch}=\frac{\De_2}{P^+}=-\frac{\tilde{\De}_2}{q^+};~~
D^{exch}=-\frac{\tilde{\De}_3}{q^+}\nn\\
&& D_1^{ann}=\frac{M_0^2}{P^+};~~
D_2^{ann}=\frac{M_0^{' 2}}{P^+};~~
D^{ann}=\frac{M_n^2}{P^+}
\leea{r17}
and from now on we use the rescaled value of the cutoff
$\la\rightarrow \la^2/P^+$.
Then the following terms contribute to the effective
electron-positron interaction \fig{1}

in the {\bf exchange channel} 
\bea
V_{\la}^{gen} &=& -e^2N_1
\left(
\frac{1}{\tilde{\De}_1}
\int_{\la}^{\infty}\frac{df_{\la'}(\De_1)}{d\la'}f_{\la'}(\De_2)d\la'+
\frac{1}{\tilde{\De}_2}
\int_{\la}^{\infty}\frac{df_{\la'}(\De_2)}{d\la'}f_{\la'}(\De_1)d\la'
\right)\nn\\
V_{\la}^{PT} &=& -e^2N_1
\frac{1}{\tilde{\De}_3}f_{\la}(\De_1)f_{\la}(\De_2)\nn\\
V_{\la}^{inst} &=& -e^2<\ga^{\mu}\ga^{\nu}>_{exch}\eta_{\mu}\eta_{\nu}
\frac{1}{q^{+2}}
\leea{r18}
in the {\bf annihilation channel} 
\bea
V_{\la}^{gen} &=& e^2N_2
\left(
\frac{1}{M_0^2}
\int_{\la}^{\infty}\frac{df_{\la'}(M_0^2)}{d\la'}f_{\la'}(M_0^{'2})d\la'+
\frac{1}{M_0^{'2}}
\int_{\la}^{\infty}\frac{df_{\la'}(M_0^{'2})}{d\la'}f_{\la'}(M_0^2)d\la'
\right)\nn\\
V_{\la}^{PT} &=& e^2N_2
\frac{1}{M_n^2}f_{\la}(M_0^2)f_{\la}(M_0^{'2})\nn\\
V_{\la}^{inst} &=& -e^2<\ga^{\mu}\ga^{\nu}>_{ann}\eta_{\mu}\eta_{\nu}
\frac{1}{P^{+2}}
\leea{r19}
where
\bea
&& N_1=-<\ga^{\mu}\ga^{\nu}>_{exch}D_{\mu\nu}(q)\nn\\
&& N_2=-<\ga^{\mu}\ga^{\nu}>_{ann}D_{\mu\nu}(P)
\leea{r20}
$q=p'_1-p_1$ is the exchanged photon momentum, with $q^-=\frac{q^{\perp 2}}{q^+}$;
and $P=p_1+p_2$ is the total momentum. 
The matrix elements 
of the effective interaction: current-current terms $N_1, N_2$
and $<\ga^{\mu}\ga^{\nu}>\eta_{\mu}\eta_{\nu}$ in both channels,
are calculated according to the rules of light-front perturbation
theory as formulated by Brodsky, Lepage \cite{LeBr}.
(see Appendices A and B for the exchange and annihilation channels, 
respectively).
The energy denominators in \eq{r18} \eq{r19} read
\bea
\tilde{\De}_1 &=& \frac{(x\vec{k}'_{\bot}-x'\vec{k}_{\bot})^2
+m^2(x-x')^2}{xx'}\; ; \qquad
 \tilde{\De}_2=\tilde{\De}_1|_{x\rightarrow(1-x),x'\rightarrow(1-x')} \nn \\
&&\De_1=\frac{\tilde{\De}_1}{x-x'} \; ; \qquad
 \De_2=\frac{\tilde{\De}_2}{x-x'} \nn \\
\tilde{\De}_3 &=& (\vec{k}_{\bot}-\vec{k}'_{\bot})^2
+\frac{(x-x')}{2}\left(\vec{k}_{\bot}^2\left(\frac{1}{1-x}-\frac{1}{x}\right)
-\vec{k}_{\bot}^{'2}\left( \frac{1}{1-x'}-\frac{1}{x'} \right)\right)\nn\\
&+& m^2\frac{(x-x')^2}{2}\left(\frac{1}{xx'}+\frac{1}{(1-x)(1-x')}\right)
+|x-x'|\left( \frac{1}{2}(M_0^2+M_0^{'2})-M_n^2 \right) \nn \\
&&M_0^2=\frac{k_{\bot}^2+m^2}{x(1-x)} \; ; \qquad
 M_0^{'2}=\frac{k_{\bot}^{'2}+m^2}{x'(1-x')}\nn\\ 
&&P^-=\frac{(P^{\bot})^2+M_n^2}{P^+} \; ; \qquad P=(P^+,P^{\bot}) 
\leea{r21}
Note $\tilde{\De}_1,~\tilde{\De}_2,~\tilde{\De}_3$ are positive defined.

The exchange channel brings the dominant contribution to the mass spectrum.
In what follows we focus on the effective interaction
in the exchange channel.

One can simplify the current-current term in the exchange channel, $N_1$.
Note, that for the vector $[q_{\mu}-(p'_{1\mu}-p_{1\mu})]$
$'+'$ and $'\perp'$ components vanish by momentum conservation,
i.e. it is proportional to the null vector $\eta_{\mu}$.
Therefore one can represent
\bea
&& q_\mu=p'_{1\mu}-p_{1\mu}-\eta_\mu \frac{D_1}{2}\nn\\
&& q_\nu=p_{2\nu}-p'_{2\nu}-\eta_\nu \frac{D_2}{2}
\leea{r21a}
Making use of the Dirac equation, one has
\bea
\hspace{-1cm}
&& \overline u(p_1,\la_1)\ga^\mu u(p'_1,\la'_1)
\overline v(p'_2,\la'_2)\ga^\nu v(p_2,\la_2)
\times (\eta_\mu q_\nu+\eta_\nu q_\mu)\nn\\
\hspace{-1cm}
&& = -\overline u(p_1,\la_1)\ga^\mu u(p'_1,\la'_1)
\overline v(p'_2,\la'_2)\ga^\nu v(p_2,\la_2)
\eta_\mu\eta_\nu \left( \frac{D_1+D_2}{2}\right)
\leea{r21b}
Then the current-current term in the exhcnage channel can be written
\bea
&& N_1=-<\ga^{\mu}\ga^{\nu}>D_{\mu\nu}(q)\rightarrow
<\ga^{\mu}\ga^{\nu}>g_{\mu\nu}
-<\ga^{\mu}\ga^{\nu}>\eta_{\mu}\eta_{\nu}
\frac{\tilde{\De}_1+\tilde{\De}_2}{2q^{+2}}
\leea{r22}
we omit the label 'exch'- exchange channel.

Below we use the approximation
\bea
&& M_n^2=\frac{1}{2}(M_0^2+M_0^{'2})
\leea{r23}
that simplifies the calculations. In this approximation
the following holds
\bea
&& \tilde{\De}=\frac{\tilde{\De}_1+\tilde{\De}_2}{2}
\leea{r24}
where now the energy denominator for the perturbative photon 
exchange reads
\bea
\tilde{\De} &=& \tilde{\De}_3(M_n^2=\frac{1}{2}(M_0^2+M_0^{'2})\nn\\
&=& (\vec{k}_{\bot}-\vec{k}'_{\bot})^2
+\frac{(x-x')}{2}\left(\vec{k}_{\bot}^2\left(\frac{1}{1-x}-
\frac{1}{x}\right)
-\vec{k}_{\bot}^{'2}\left( \frac{1}{1-x'}-\frac{1}{x'} \right)\right)\nn\\
&+& m^2\frac{(x-x')^2}{2}\left(\frac{1}{xx'}+\frac{1}{(1-x)(1-x')}\right)
\leea{r25}

Combining all the terms \eq{r18} together 
one has for the effective interaction in the exchange channel
\bea
V_{LC}^{eff}(\la) &=& V_{\la}^{gen}+V_{\la}^{PT}+V_{\la}^{inst}\nn\\
&=& -e^2<\ga^{\mu}\ga^{\nu}>g_{\mu\nu}
\left(\frac{\Theta_{1\la}}{\tilde{\De}_1}+\frac{\Theta_{2\la}}
{\tilde{\De}_2}
+\frac{f_{1\la}f_{2\la}}{\tilde{\De}}\right)\nn\\
&& -e^2<\ga^{\mu}\ga^{\nu}>\eta_{\mu}\eta_{\nu}
\frac{1}{2q^{+2}}(\tilde{\De}_1-\tilde{\De}_2)
\left(\frac{\Theta_{1\la}}{\tilde{\De}_1}
-\frac{\Theta_{2\la}}{\tilde{\De}_2}\right)
\leea{r26}
where we have introduced
\bea
&& \Theta_{1\la}=\int_{\la}^{\infty}\frac{df_{\la'}(\De_1)}{d\la'}
f_{\la'}(\De_2)d\la'\nn\\
&& \Theta_{2\la}=\int_{\la}^{\infty}\frac{df_{\la'}(\De_2)}{d\la'}
f_{\la'}(\De_1)d\la'\nn\\
&& f_{1\la}=f_{\la}(\De_1);~~f_{2\la}=f_{\la}(\De_2)
\leea{r27}
 and one has
\bea
&& \Theta_{1\la}+\Theta_{2\la}=1-f_{1\la}f_{2\la}
\leea{r28}
Remind, that $f_{\la}(\De)$ is the similarity function specified below.

Instantaneous exchange with the energy below the cutoff
$e^2<\ga^{\mu}\ga^{\nu}>\eta_{\mu}\eta_{\nu}
\frac{1}{q^{+2}}f_{1\la}f_{2\la}$ is canceled in the effective
interaction $V_{LC}^{eff}(\la)$ \eq{r26}.

In the \eq{r26} the "$\eta_{\mu}\eta_{\nu}$" term is spin-independent,
and it is at least one power of momenta higher than the leading
spin-independent piece in the "$g_{\mu\nu}$" term. This form 
of the effective interaction \eq{r26} is convenient to separate the structures:
"$\eta_{\mu}\eta_{\nu}$" term, containing the collinear singularity
when $x\sim x'$, from the $"g_{\mu\nu}"$ term.

The initial value of the effective interaction is given
at the bare cutoff $\La$ 
\bea
&& V_{LC}^{eff}(\la=\La\rightarrow\infty)
=V_{\la=\La\rightarrow\infty}^{PT}+V^{inst}
=-e^2<\ga^{\mu}\ga^{\nu}>g_{\mu\nu}
\frac{1}{\tilde{\De}}
\leea{r29}
This is the result of light-front perturbative theory \cite{LeBr}.
Note, that though $V^{inst}$ is singular as $x\sim x'$,
the interaction $V_{\la=\La\rightarrow\infty}^{PT}+V^{inst}$
is free of collinear singularity.

The resulting value of the effective interaction
$V_{LC}^{eff}$ is defined at $\la\rightarrow 0$
(except maybe for the point of Coulomb singularity 
$\vec{q}(q_z,\vec{q}_{\perp})=0$, i.e. 
$(x=x',\vec{k}_{\perp}=\vec{k}^{'}_{\perp}))$
\bea
V_{LC}^{eff} &=& V_{LC}^{eff}(\la\rightarrow 0 )
=V_{\la\rightarrow 0}^{gen}+V^{inst}\nn\\
&=& -e^2<\ga^{\mu}\ga^{\nu}>g_{\mu\nu}
\left(\frac{\Theta_{1}}{\tilde{\De}_1}+\frac{\Theta_{2}}{\tilde{\De}_2}
\right)\nn\\
&& -e^2<\ga^{\mu}\ga^{\nu}>\eta_{\mu}\eta_{\nu}
\frac{1}{2q^{+2}}(\tilde{\De}_1-\tilde{\De}_2)
\left(\frac{\Theta_{1}}{\tilde{\De}_1}
-\frac{\Theta_{2}}{\tilde{\De}_2}\right)
\leea{r30}
where
\bea
&& \Theta_1=\Theta_{1\la}|_{\la=0}\nn\\
&& \Theta_2=\Theta_{2\la}|_{\la=0}
\leea{r31}

Estimate the difference between the initial and final values 
of the effective interaction. Introduce 
$\tilde{\De}_1-\tilde{\De}_2=2\de$, then together with \eq{r24}
one has
\bea
&& \tilde{\De}_1=\tilde{\De}+\de;~~\tilde{\De}_2=\tilde{\De}-\de\nn\\
&& \de=\frac{(x-x')}{2}\left(-\frac{k_{\perp}^2}{x(1-x)}
+\frac{k_{\perp}^{'2}}{x'(1-x')}
+m^2(x-x')(\frac{1}{xx'}-\frac{1}{(1-x)(1-x')})\right)
\leea{r32}
where usually (in the nonrelativistic case and
in the collinear limit $x\sim x'$)
holds $|\de|<<\tilde{\De}$.
For the effective interaction one has
\bea
V_{LC}^{eff} &=& -e^2<\ga^{\mu}\ga^{\nu}>g_{\mu\nu}\frac{1}{\tilde{\De}}
\left(1-(\Theta_1-\Theta_2)\frac{\de}{\tilde{\De}}
+O(\frac{\de^2}{\tilde{\De}^2})\right)\nn\\
&& -e^2<\ga^{\mu}\ga^{\nu}>\eta_{\mu}\eta_{\nu}
\frac{1}{q^{+2}}\left((\Theta_1-\Theta_2)\frac{\de}{\tilde{\De}}
+O(\frac{\de^2}{\tilde{\De}^2})\right)\nn\\
&=& V^{(0)}+\sum_i\De V^{(i)}_{g_{\mu\nu}}
+\sum_i\De V^{(i)}_{\eta_{\mu}\eta_{\nu}}
\leea{r33}
index above in $\De V^{(i)}$ shows the order of expansion
with respect to $(\de/\tilde{\De})$.
The leading order in \eq{r33} is given by the result of perturbation theory
\bea
&& V^{(0)}=-e^2<\ga^{\mu}\ga^{\nu}>g_{\mu\nu}
\frac{1}{\tilde{\De}}=V_{\la=\La\rightarrow\infty}^{PT}+V^{inst}
\leea{r35}
We change the variables, which defines $p_z$ :~
$(x,\vec{k}_{\perp})\rightarrow \vec{p}(p_z,\vec{k}_{\perp})$,~
$\vec{p}$ is the three momentum in the center of mass frame
\bea
&& x=\frac{1}{2}\left(1+\frac{p_z}{\sqrt{\vec{p}^2+m^2}}\right)
\leea{r34}
The term \eq{r35} gives in the leading order 
of nonrelativistic approximation
$|\vec{p}|/m<<1$ the $3$-dimensional Coulomb interaction
\bea
&& V^{(0)}\approx -\frac{16e^2m^2}{\vec{q}^2}
\leea{r36}
where $\vec{q}(q_z,\vec{q}_{\perp})=\vec{p'}-\vec{p}$ 
is the exchanged momentum.

It was shown in \cite{JoPeGl},\cite{BrPe}, that the nonrelativistic
expansion of the term \eq{r35} to the~ second order 
$O((\frac{\vec{p}}{m})^2)$ gives rise to the correct spin splittings
of positronium and the rotational invariance
(due to the degeneracy of triplet states) is restored.

Corrections $\De V_{g_{\mu\nu}};\De V_{\eta_{\mu}\eta_{\nu}}$ 
arise due to the unitary transformation,
i.e. the corrections due 
to the energy denominators in the "$g_{\mu\nu}$" term
and the "$\eta_{\mu}\eta_{\nu}$" term.
Estimate the first order corrections 
$\De V^{(1)}_{g_{\mu\nu}};\De V^{(1)}_{\eta_{\mu}\eta_{\nu}}$ 
$O(\frac{\de}{\tilde{\De}})$
in the nonrelativistic case. We choose for this purpose
the explicit form of the similarity function  
\bea
&& f_{\la}(\De)={\rm e}^{-\frac{\De^2}{\la^4}}
\leea{r37}
Then one has
\bea
&& \Theta_1=\frac{\tilde{\De}_1^2}{\tilde{\De}_1^2+\tilde{\De}_2^2}
\leea{}
and
\bea
&& \hspace{-2cm}
V_{LC}^{eff}=-e^2<\ga^{\mu}\ga^{\nu}>g_{\mu\nu}
\frac{\tilde{\De}_1+\tilde{\De}_2}{\tilde{\De}_1^2+\tilde{\De}_2^2}
-e^2<\ga^{\mu}\ga^{\nu}>\eta_{\mu}\eta_{\nu}\frac{1}{2q^{+2}}
\frac{(\tilde{\De}_1-\tilde{\De}_2)^2}{\tilde{\De}_1^2+\tilde{\De}_2^2}
\leea{r38}
The series \eq{r33} for the effective interaction then reads
\bea
V_{LC}^{eff} &=& -e^2<\ga^{\mu}\ga^{\nu}>g_{\mu\nu}\frac{1}{\tilde{\De}}
\left(1-\frac{\de^2}{\tilde{\De}^2}
+O(\frac{\de^3}{\tilde{\De}^3})\right)\nn\\
&& -e^2<\ga^{\mu}\ga^{\nu}>\eta_{\mu}\eta_{\nu}
\frac{1}{q^{+2}}\left(\frac{\de^2}{\tilde{\De}^2}
+O(\frac{\de^3}{\tilde{\De}^3})\right)
\leea{r39}
where $\de\approx -\frac{q_z}{m}(\vec{q},\vec{p}+\vec{p'})$.

The corrections are given correspondingly
\bea
&& \De V^{(1)}_{g_{\mu\nu}}=
e^2<\ga^{\mu}\ga^{\nu}>g_{\mu\nu}\frac{1}{\tilde{\De}}
\frac{\de^2}{\tilde{\De}^2}\approx
\frac{16e^2m^2}{\vec{q}^2}
\left(\frac{q_z(\vec{q},\vec{p}+\vec{p'})}{m\vec{q}^2}\right)^2\nn\\
&& \De V^{(1)}_{\eta_{\mu}\eta_{\nu}}=
e^2<\ga^{\mu}\ga^{\nu}>\eta_{\mu}\eta_{\nu}
\frac{1}{q^{+2}}\frac{\de^2}{\tilde{\De}^2}\approx
\frac{16e^2m^2}{q_z^2}
\left(\frac{q_z(\vec{q},\vec{p}+\vec{p'})}{m\vec{q}^2}\right)^2
\leea{r40}
where $x-x'\approx \frac{q_z}{2m}$. 

The corrections due to energy denominators in the 
$"g_{\mu\nu}"$ term are spin-independent and do not affect
the spin-dependent interactions.

The correction from
$"\eta_{\mu}\eta_{\nu}"$ term is of the order $e^2q^0$.
Quite generally the interaction given by the light-front perturbation theory
$V_{\la=\La\rightarrow\infty}^{PT}+V^{inst}$ and the effective interaction
generated by the unitary transformation $V_{\la\rightarrow 0}^{gen}+V^{inst}$
have both a leading Coulomb behavior \eq{r36}, but they differ
by spin-independent $"\eta_{\mu}\eta_{\nu}"$ term in the order $e^2q^0$,
which contributes in the first order bound state perturbation
theory to the mass in the order $\alpha^4$.
In the order of fine
structure splitting $\alpha^4$ also terms of order $e^4q^{-1}$
and $e^6q^{-2}$ are important \cite{GuWe}. 
We  expect that the structure $"\eta_{\mu}\eta_{\nu}"$ in order
$e^2q^0$ will be compensated in the mass spectrum 
(in its spin-independent part) 
by the corresponding terms in the order $(e^4)$ and $(e^6)$. 
\footnote{For the similarity function
$f_{\la}(\De)=\theta(\la^2-|\De|)$ the corrections are
\bea
&& \De V^{(1)}_{g_{\mu\nu}}=
e^2<\ga^{\mu}\ga^{\nu}>g_{\mu\nu}\frac{1}{\tilde{\De}}
\frac{|\de|}{\tilde{\De}}\approx
\frac{16e^2m^2}{\vec{q}^2}
\frac{|q_z||\vec{q}(\vec{p}+\vec{p'})|}{m\vec{q}^2}\nn\\
&& \De V^{(1)}_{\eta_{\mu}\eta_{\nu}}=
e^2<\ga^{\mu}\ga^{\nu}>\eta_{\mu}\eta_{\nu}
\frac{1}{q^{+2}}\frac{|\de|}{\tilde{\De}}\approx
\frac{16e^2m^2}{q_z^2}
\frac{|q_z||\vec{q}(\vec{p}+\vec{p'})|}{m\vec{q}^2}
\leea{r40}
In this case the "$\eta_{\mu}\eta_{\nu}$" term is of the order $e^2q^{-1}$
and contributes to the mass in the order $\alpha^3$.
One needs to calculate the effective interaction up to the term $e^4q^{-2}$
to cancel this contribution from "$\eta_{\mu}\eta_{\nu}$" term in the mass.

In the case of $f_{\la}(\De)={\rm e}^{-\frac{|\De|}{\la^2}}$,
(see the main text)
\bea
&& \De V_{g_{\mu\nu}}=0,~~\De V_{\eta_{\mu}\eta_{\nu}}=0.
\leea{} 
}

It is argued in \cite{BrPe}, that though $"\eta_{\mu}\eta_{\nu}"$ term
does not affect spin-spin and tensor interactions,
but it may influence in the second order
bound state perturbation theory the spin-orbit.

We remind also that $"\eta_{\mu}\eta_{\nu}"$ term 
in the effective interaction can have the singular behavior 
in the collinear limit as $x\sim x'$ (Appendix C).

One can generate the result of perturbation theory
by the flow equations \cite{GuWe}, if one chooses for 
the similarity function
\bea
&& f_{\la}(\De)={\rm e}^{-\frac{|\De|}{\la^2}}
\leea{r41}
Then $\Theta$-factor \eq{r31} is 
$\Theta_1=\frac{\tilde{\De}_1}{\tilde{\De}_1+\tilde{\De}_2}$;
$"\eta_{\mu}\eta_{\nu}"$ term in \eq{r30} vanishes and one has for 
the effective interaction
\bea
&& V_{LC}^{eff}=-e^2<\ga^{\mu}\ga^{\nu}>g_{\mu\nu}\frac{1}{\tilde{\De}}
\leea{r42}
This is true with the approximation done before \eqs{r23}{r24}.

Other choices of the similarity functions $f_{\la}(\De)$ are possible
(\eqs{f22}{f23}, see also Appendix C). 
We expect that the mass spectrum stay intact
with respect to different similarity functions.

In order to introduce the spectroscopic notation for positronium
mass spectrum we integrate out the angular degree of freedom
($\varphi$) by substituting it with the discrete quantum number
$J_z=n$, $n\in {\bf Z}$
(actually for the annihilation channel only $|J_z|\leq 1$ is possible)
\bea
&& \hspace{-2cm} <x,k_{\perp};J_z,\la_1,\la_2|\tilde{V}_{LC}^{eff}|
x',k'_{\perp};J'_z,\la'_1,\la'_2>\nn\\
&=& \frac{1}{2\pi}\int_0^{2\pi}d\varphi{\rm e}^{-iL_z\varphi}
\int_0^{2\pi}d\varphi'{\rm e}^{iL'_z\varphi'}(-\frac{1}{2(2\pi)^3})
<x,k_{\perp},\varphi;\la_1,\la_2|V_{LC}^{eff}|
x',k'_{\perp},\varphi';\la'_1,\la'_2>\nn\\
&&
\leea{r43}
where $L_z=J_z-S_z$; $S_z=\frac{\la_1}{2}+\frac{\la_2}{2}$ and the states
can be classified (strictly speaking only for rotationally invariant
systems) according to their quantum numbers of total angular momentum $J$,
orbit angular momentum $L$, and total spin $S$. 
It should be noted
that the definition of angular momentum operators in light-front
dynamics is problematic because they include the interaction.

The general matrix elements for the effective interaction
depending on the angles $\varphi,\varphi'$
(actually on the difference $(\varphi -\varphi')$)
$<x,k_{\perp},\varphi;\la_1,\la_2|V_{LC}^{eff}|
x',k'_{\perp},\varphi';\la'_1,\la'_2>$
and also the matrix elements of the effective interaction
after the angular integration
for the total momentum $J_z$
$<x,k_{\perp};J_z,\la_1,\la_2|\tilde{V}_{LC}^{eff}|
x',k'_{\perp};J'_z,\la'_1,\la'_2>$
in the exchange and annihilation channels are given
in Appendices A and B, respectively.

To perform the angular integration \eq{r43} analytically
we choose for the similarity function \eq{f13}
\bea
&& f_{\la}(\De)=u_{\la}(\De)=\theta(\la^2-|\De|)
\leea{r44}
where for simplicity we use the sharp cutoff function $u_{\la}(\De)$.
Then the effective interaction reads
\bea
V_{LC}^{eff} &=&
   -e^2<\ga^{\mu}\ga^{\nu}>g_{\mu\nu}
\left(\frac{\theta(\tilde{\De}_1-\tilde{\De}_2)}{\tilde{\De}_1}
+\frac{\theta(\tilde{\De}_2-\tilde{\De}_1)}{\tilde{\De}_2}
\right)\nn\\
&& -e^2<\ga^{\mu}\ga^{\nu}>\eta_{\mu}\eta_{\nu}
\frac{1}{2q^{+2}}|\tilde{\De}_1-\tilde{\De}_2|
\left(\frac{\theta(\tilde{\De}_1-\tilde{\De}_2)}{\tilde{\De}_1}
+\frac{\theta(\tilde{\De}_2-\tilde{\De}_1)}{\tilde{\De}_2}
\right)
\leea{r45}
where $\theta(x)-\theta(-x)=sign(x)$.

We proceed now to solve for the positronium spectrum 
in all sectors of $J_z$. For this purpose we formulate 
the light-front integral equation \eq{r4} in the form
where the integral kernel is given by the effective interaction
for the total momentum $J_z$ \eq{r43}. After the change 
of variables \eq{r34}
$(\vec{k}_{\perp};x)=(k_{\perp},\varphi;x)\rightarrow
\vec{p}=(\vec{k}_{\perp},p_z)=
(\mu\sin\theta\cos\varphi,\mu\sin\theta\sin\varphi,\mu\cos\theta)$
\bea
&& x=\frac{1}{2}\left(1+\frac{\mu\cos\theta}{\sqrt{\mu^2+m^2}}\right)
\leea{r46}
where the Jacobian reads
\bea
&& J=\frac{\mu^2}{2}
\frac{m^2+\mu^2(1-\cos^2\theta)}{(m^2+\mu^2)^{3/2}}\sin\theta
\leea{r47}
one has the following integral equation
\bea
&& \hspace{-2cm} 
(M_n^2-4(m^2+\mu^2))\tilde{\psi}_n(\mu,\cos\theta;J_z,\la_1,\la_2)
+\sum_{J'_z,\la'_1,\la'_2}\int_{D}d\mu'\int_{-1}^{+1}d\cos\theta'
\frac{\mu^{'2}}{2}
\frac{m^2+\mu^{'2}(1-\cos^2\theta')}{(m^2+\mu^{'2})^{3/2}}\nn\\
&& \hspace{0.5cm}
\times <\mu,\cos\theta;J_z,\la_1,\la_2|\tilde{V}_{LC}^{eff}|
\mu',\cos\theta';J'_z,\la'_1,\la'_2>
\tilde{\psi}_n(\mu',\cos\theta';J'_z,\la'_1,\la'_2)=0\nn\\
&&
\leea{r48}
The integration domain $D$ \eq{r6} is given now by 
$\mu\in [0;\frac{\La}{2}]$.
Neither $L_z$ nor $S_z$ are good quantum numbers; therefore
we set $L_z=J_z-S_z$.

The wave function is normalized
\bea
&& \sum_{J_z,\la_1,\la_2}\int d\mu~d\cos\theta
\tilde{\psi}_n^{*}(\mu,\cos\theta;J_z,\la_1,\la_2)
\tilde{\psi}_{n'}(\mu,\cos\theta;J_z,\la_1,\la_2)
=\de_{nn'}
\leea{r49}
where $n$ labels all quantum numbers.

The integral equation \eq{r48} may be used for the calculations
of positronium mass spectrum numerically.

\section{Conclusions}

We have considered the particle number conserving part 
of the effective QED Hamiltonian, generated in second order
in the coupling by unitary transformation.
The new generated interaction between electron and positron
has two terms of different structure.

The first term has spin-independent as well as spin-dependent parts.
The spin-independent part, combined with the instantaneous
exchange interaction, leads to the Coulomb interaction.
The light-front spin-dependent part after 
a simple unitary transformation to rotate the spins
gives the familiar Breit-Fermi spin-spin and tensor interactions
\cite{BrPe}, that lead to the correct spin splittings 
of positronium and rotational invariance is restored \cite{JoPeGl}.

These properties of the first term do not depend on 
the explicit form of the unitary transformation performed.

The second term is spin-independent and do depend
on the unitary transformation. In the first order bound state
perturbation theory it contributes the spin-independent
part to the mass in order of fine structure.
We expect, that this contribution in mass is cancelled
by the terms from higher orders in the unitary transformation.
The second term  may also influence
the spin-orbit in the second order bound state 
perturbation theory \cite{BrPe}.

\paragraph{Acknowledgments}
The author would like to thank Prof. F.Wegner and
Dr. Martina Brisudova for usefull discussions.

\newpage
\appendix

\section{Matrix elements of the effective interaction.Exchange channel}
\label{A}

In the Appendices A and B we follow the scheme of the work \cite{TrPa}
to calculate the matrix elements of the effective interaction
in the exchange and annihilation channels, respectively.
Here, we list the general, angle-dependent matrix elements
defining the effective interaction in the exchange channel 
\eqs{r26}{r30} (part I) 
and the corresponding matrix elements
of the effective interaction for arbitrary $J_z$, after integrating out
the angles \eq{r43} (part II). The whole is given for the similarity
function $f_{\la}(\De)=u_{\la}(\De)=\theta(\la^2-|\De|)$ \eq{f13}
with the sharp cutoff.
The effective interaction generated by the unitary transformation
in the exchange channel reads \eq{r45}
\bea
V_{LC}^{eff}
&=& -e^2<\ga^{\mu}\ga^{\nu}>g_{\mu\nu}
\left(\frac{\theta(a_1-a_2)}{\tilde{\De}_1}
+\frac{\theta(a_2-a_1)}{\tilde{\De}_2}
\right)\nn\\
&& -e^2<\ga^{\mu}\ga^{\nu}>\eta_{\mu}\eta_{\nu}\frac{1}{2q^{+2}}
\left(\frac{(a_1-a_2)\theta(a_1-a_2)}{\tilde{\De}_1}
+\frac{(a_2-a_1)\theta(a_2-a_1)}{\tilde{\De}_2}
\right)\nn\\
&=& -e^2<\ga^{\mu}\ga^{\nu}>g_{\mu\nu}
\left(\frac{\theta(a_1-a_2)}{\tilde{\De}_1}
+\frac{\theta(a_2-a_1)}{\tilde{\De}_2}
\right)\nn\\
&& -e^2<\ga^{\mu}\ga^{\nu}>\eta_{\mu}\eta_{\nu}\frac{1}{2q^{+2}}
|a_1-a_2|
\left(\frac{\theta(a_1-a_2)}{\tilde{\De}_1}
+\frac{\theta(a_2-a_1)}{\tilde{\De}_2}
\right)
\leea{a1}
where \fig{1}
\bea
&& <\ga^{\mu}\ga^{\nu}>|_{exch}
=\frac{\bar{u}(p_1,\la_1)}{\sqrt{p_1^{+}}}\ga^{\mu}
\frac{u(p_1^{'},\la_1^{'})}{\sqrt{p_1^{' +}}}
\frac{\bar{v}(p_2^{'},\la_2^{'})}{\sqrt{p_2^{' +}}}\ga^{\nu}
\frac{v(p_2,\la_2)}{\sqrt{p_2^{+}}}
P^{+ 2}
\leea{a1a}
$q=p'_1-p_1$ is the momentum transfer. One has in \eq{a1}
\bea
&& \tilde{\De}_1=a_1-2k_{\perp}k_{\perp}^{'}\cos(\varphi-\varphi^{'})\nn\\
&& \tilde{\De}_2=a_2-2k_{\perp}k_{\perp}^{'}\cos(\varphi-\varphi^{'})\nn\\
&& \tilde{\De}=a-2k_{\perp}k_{\perp}^{'}\cos(\varphi-\varphi^{'})\nn\\
&& \vec{k}_{\perp}=k_{\perp}(\cos\varphi,\sin\varphi)
\leea{}
and
\bea
a_1 &=& \frac{x'}{x}k_{\perp}^2+\frac{x}{x'}k_{\perp}^{'2}
+m^2\frac{(x-x')^2}{xx'}\nn\\
&=& k_{\perp}^2+k_{\perp}^{'2}
+(x-x')\left(k_{\perp}^2(-\frac{1}{x})-k_{\perp}^{'2}(-\frac{1}{x'})\right)
+m^2\frac{(x-x')^2}{xx'}\nn\\
a_2 &=& \frac{1-x'}{1-x}k_{\perp}^2+\frac{1-x}{1-x'}k_{\perp}^{'2}
+m^2\frac{(x-x')^2}{(1-x)(1-x')}\nn\\
&=& k_{\perp}^2+k_{\perp}^{'2}
+(x-x')\left(k_{\perp}^2\frac{1}{1-x}-k_{\perp}^{'2}\frac{1}{1-x'}\right)
+m^2\frac{(x-x')^2}{(1-x)(1-x')}\nn\\
a &=& k_{\perp}^2+k_{\perp}^{'2}
+\frac{(x-x')}{2}\left(k_{\perp}^2(\frac{1}{1-x}-\frac{1}{x})
-k_{\perp}^{'2}(\frac{1}{1-x'}-\frac{1}{x'})\right)\nn\\
&+& m^2\frac{(x-x')^2}{2}\left(\frac{1}{xx'}
+\frac{1}{(1-x)(1-x')}\right)\nn\\
a &=& \frac{1}{2}(a_1+a_2)
\leea{a3}
The energy denominator ($\tilde{\De}$ and $a$ corresponding)
in the case of perturbative theory is given for completeness.

It is useful to display the matrix elements of the effective interaction
in the form of tables. The matrix elements depend on the one hand
on the momenta of the electron and positron, respectively, and on the other
hand on their helicities before and after the interaction.
The dependence on the helicities occur during the calculation
of these functions
$E(x,\vec{k}_{\perp};\la_1,\la_2|x',\vec{k}'_{\perp};\la'_1,\la'_2)$
in part I and 
$G(x,k_{\perp};\la_1,\la_2|x',k'_{\perp};\la'_1,\la'_2)$ in part II
as different Kronecker deltas \cite{LeBr}.
These functions are displayed in the form of helicity tables.
We use the following notation for the elements of the tables
\bea
&& F_i(1,2)~\rightarrow ~E_i(x,\vec{k}_{\perp};x',\vec{k}'_{\perp});~
G_i(x,k_{\perp};x',k'_{\perp})
\leea{a4}
Also we have used in both cases for the permutation of particle and
anti-particle
\bea
&&F_3^{*}(x,\vec{k}_{\perp};x',\vec{k}'_{\perp})
=F_3(1-x,-\vec{k}_{\perp};1-x',-\vec{k}'_{\perp})
\leea{a5}
one has the corresponding for the elements of arbitrary $J_z$;
in the case when the function additionally depends
on the component of the total angular momentum $J_z=n$
we have introduced
\bea
&& \tilde{F}_i(n)=F_i(-n)
\leea{a6}

\subsection{The general helicity table.}

To calculate the matrix elements of the effective interaction
in the exchange channel we use the matrix elements of the Dirac spinors
listed in Table $1$ \cite{LeBr}. 
Also the following holds
$\bar{v}_{\la'}(p)\ga^{\alpha}v_{\la}(q)
=\bar{u}_{\la}(q)\ga^{\alpha}u_{\la'}(p)$.

\begin{table}[htb]
\centerline{
\begin{tabular}{|c||c|}
\hline
\parbox{1.5cm}{ \[\cal M\] } & 
\parbox{12cm}{
\[
\frac{1}{\sqrt{k^+k^{'+}}}
\bar{u}(k',\lambda') {\cal M} u(k,\lambda)
\]
}
\\\hline\hline
$\gamma^+$ & \parbox{12cm}{
\[
\hspace{3cm}
2\delta^{\lambda}_{\lambda'}
\]}
\\\hline
$\gamma^-$ &
\parbox{12cm}{
\[
\hspace{-1cm}
\frac{2}{k^+k^{'+}}\left[\left(m^2+
k_{\perp}k'_{\perp}e^{+i\lambda(\varphi-\varphi')}\right)
\delta^{\lambda}_{\lambda'}
-m\lambda\left(k'_{\perp}e^{+i\lambda \varphi'}-
k_{\perp} e^{+i\lambda\varphi}\right)
\delta^{\lambda}_{-\lambda'}\right]
\]}
\\\hline
$\gamma_{\perp}^1$ & 
\parbox{12cm}{
\[
\left(\frac{k'_{\perp}}{k^{'+}}e^{-i\lambda\varphi'}+
\frac{k_{\perp}}{k^+}e^{+i\lambda\varphi}
\right)\delta^{\lambda}_{\lambda'}
+m\lambda\left(\frac{1}{k^{'+}}-\frac{1}{k^+}
\right)\delta^{\lambda}_{-\lambda'}
\]}
\\\hline
$\gamma_{\perp}^2$ &
\parbox{12cm}{
\[ 
i\lambda\left(\frac{k'_{\perp}}{k^{'+}}e^{-i\lambda\varphi'}-
\frac{k_{\perp}}{k^+}e^{+i\lambda\varphi}
\right)\delta^{\lambda}_{\lambda'}
+im\left(\frac{1}{k^{'+}}-\frac{1}{k^+}
\right)\delta^{\lambda}_{-\lambda'}
\]}\\
\hline
\end{tabular}
}
\caption[Matrix elements of the Dirac spinors]
{Matrix elements of the Dirac spinors.}
\end{table}

We introduce for the matrix elements entering in the effective
interaction \eq{a1}
\bea
2E^{(1)}(x,\vec{k}_{\perp};\la_1,\la_2|
x',\vec{k}_{\perp}^{'};\la_1^{'},\la_2^{'})
&=& <\ga^{\mu}\ga^{\nu}>g_{\mu\nu}=\nn\\
&=& \frac{1}{2}<\ga^+\ga^->+\frac{1}{2}<\ga^-\ga^+>
-<\ga_1^2>-<\ga_2^2>\nn\\
2E^{(2)}(x,\vec{k}_{\perp};\la_1,\la_2|
x',\vec{k}_{\perp}^{'};\la_1^{'},\la_2^{'}) 
&=& <\ga^{\mu}\ga^{\nu}>\eta_{\mu}\eta_{\nu}\frac{1}{q^{+2}}
=<\ga^+\ga^+>\frac{1}{q^{+2}}
\leea{a7}
where
\bea
&& <\ga^{\mu}\ga^{\nu}>=\frac{\bar{u}(x,\vec{k}_{\perp};\la_1)~\ga^{\mu}
~u(x',\vec{k}_{\perp}^{'};\la_1^{'})~
~\bar{v}(1-x',-\vec{k}_{\perp}^{'};\la_2^{'})~\ga^{\nu}
~v(1-x,-\vec{k}_{\perp};\la_2)}{\sqrt{xx'(1-x)(1-x')}}
\leea{a8}
These functions are displayed in the Table $2$.

\begin{table}[htb]
\centerline{
\begin{tabular}{|c||c|c|c|c|}\hline
\rule[-3mm]{0mm}{8mm}{\bf final : initial} & 
$(\lambda_1',\lambda_2')=\uparrow\uparrow$ 
& $(\lambda_1',\lambda_2')=\uparrow\downarrow$ 
& $(\lambda_1',\lambda_2')=\downarrow\uparrow$ &
$(\lambda_1',\lambda_2')=\downarrow\downarrow$ \\ \hline\hline
\rule[-3mm]{0mm}{8mm}$(\lambda_1,\lambda_2)=\uparrow\uparrow$ & $E_1(1,2)$  
& $E_3^*(1,2)$ & $E_3(1,2)$ & $0$ \\ \hline
\rule[-3mm]{0mm}{8mm}$(\lambda_1,\lambda_2)=\uparrow\downarrow$ & 
$E_3^*(2,1)$ & $E_2(1,2)$ & $E_4(1,2)$ 
& $-E_3(2,1)$ \\ \hline
\rule[-3mm]{0mm}{8mm}$(\lambda_1,\lambda_2)=\downarrow\uparrow$& $E_3(2,1)$ 
& $E_4(1,2)$ & $E_2(1,2)$  & 
$-E_3^*(2,1)$\\ \hline
\rule[-3mm]{0mm}{8mm}$(\lambda_1,\lambda_2)=\downarrow\downarrow$ & $0$ 
& $-E_3(1,2)$ & $-E_3^*(1,2)$ & 
$E_1(1,2)$\\ \hline
\end{tabular}
}
\protect\caption{General helicity table defining the effective interaction
in the exchange channel.}
\label{GeneralHelicityTable}
\end{table}

The matrix elements 
$E_i^{(n)}(1,2)=E_i^{(n)}(x,\vec{k}_{\perp};x',\vec{k}'_{\perp})$
$(n=1,2)$ are the following
\bea
E_1^{(1)}(x,\vec{k}_{\perp};x',\vec{k}_{\perp}^{'})
&=& m^2\left(\frac{1}{xx'}+\frac{1}{(1-x)(1-x')}\right)
+\frac{k_{\perp}k_{\perp}^{'}}{xx'(1-x)(1-x')}
{\rm e}^{-i(\varphi-\varphi^{'})}\nn\\
E_2^{(1)}(x,\vec{k}_{\perp};x',\vec{k}_{\perp}^{'})
&=& m^2\left(\frac{1}{xx'}+\frac{1}{(1-x)(1-x')}\right)
+k_{\perp}^2\frac{1}{x(1-x)}+k_{\perp}^{'2}\frac{1}{x'(1-x')}\nn\\
&+& k_{\perp}k_{\perp}^{'}
\left(\frac{{\rm e}^{i(\varphi-\varphi^{'})}}{xx'}
+\frac{{\rm e}^{-i(\varphi-\varphi^{'})}}{(1-x)(1-x')}\right)\nn\\
E_3^{(1)}(x,\vec{k}_{\perp};x',\vec{k}_{\perp}^{'})
&=& -\frac{m}{xx'}
\left(k_{\perp}^{'}{\rm e}^{i\varphi^{'}}
-k_{\perp}\frac{1-x'}{1-x}{\rm e}^{i\varphi}\right)\nn\\
E_4^{(1)}(x,\vec{k}_{\perp};x',\vec{k}_{\perp}^{'})
&=& -m^2\frac{(x-x')^2}{xx'(1-x)(1-x')}
\leea{a9}
and
\bea
&& E_1^{(2)}(x,\vec{k}_{\perp};x',\vec{k}_{\perp}^{'})
=E_2^{(2)}(x,\vec{k}_{\perp};x',\vec{k}_{\perp}^{'})
=\frac{2}{(x-x')^2}\nn\\
&& E_3^{(2)}(x,\vec{k}_{\perp};x',\vec{k}_{\perp}^{'})
=E_4^{(2)}(x,\vec{k}_{\perp};x',\vec{k}_{\perp}^{'})=0
\leea{a10}

\subsection{The helicity table of the exchange channel
for arbitrary $J_z$.}

Following the description given in the main text \eq{r43}
we integrate out the angles in the effective interaction 
in the exchange channel \eqs{r45}{a1}.
For the matrix elements of the effective interaction
for an arbitrary $J_z=n$ with $n\in {\bf Z}$~~~
$G(x,k_{\perp};\la_1,\la_2|x',k'_{\perp};\la'_1,\la'_2)=
<x,k_{\perp};J_z,\la_1,\la_2|\tilde{V}_{LC}^{eff}|
x',k'_{\perp};J'_z,\la'_1,\la'_2>|_{exch}$ in the exchange channel
one obtains the helicity Table $3$.

\begin{table}[htb]
\centerline{
\begin{tabular}{|c||c|c|c|c|}\hline
\rule[-3mm]{0mm}{8mm}{\bf final : initial} 
&$(\lambda'_1,\lambda'_2)=\uparrow\uparrow$
&$(\lambda'_1,\lambda'_2)=\uparrow\downarrow$
&$(\lambda'_1,\lambda'_2)=\downarrow\uparrow$
&$(\lambda'_1,\lambda'_2)=\downarrow\downarrow$\\\hline\hline
\rule[-3mm]{0mm}{8mm}$(\lambda_1,\lambda_2)=\uparrow\uparrow$
&$G_1(1,2)$&$G_3^*(1,2)$&$G_3(1,2)$&$0$\\\hline
\rule[-3mm]{0mm}{8mm}$(\lambda_1,\lambda_2)=\uparrow\downarrow$
&$G_3^*(2,1)$&$G_2(1,2)$&$G_4(1,2)$&$-\tilde{G}_3(2,1)$\\\hline
\rule[-3mm]{0mm}{8mm}$(\lambda_1,\lambda_2)=\downarrow\uparrow$
&$G_3(2,1)$&$G_4(1,2)$&$\tilde{G}_2(1,2)$&$-\tilde{G}_3^*(2,1)$\\\hline
\rule[-3mm]{0mm}{8mm}$(\lambda_1,\lambda_2)=\downarrow\downarrow$&$0$&$-
\tilde{G}_3(1,2)$&$-\tilde{G}_3^*(1,2)$&
$\tilde{G}_1(1,2)$\\
\hline
\end{tabular}
}
\protect\caption[Helicity table of the effective interaction
in the exchange channel for arbitrary $J_z = \pm n$, $x>x'$.]
{\protect\label{HelicityTableJz}Helicity table of the effective interaction
for $J_z = \pm n$, $x>x'$.}
\end{table}
\vspace{0.5cm}

Here, the functions $G_i(1,2)=G_i(x,k_{\perp};x',k'_{\perp})$
are given
\bea
G_1(x,k_{\perp};x',k_{\perp}^{'})&=& 
m^2\left(\frac{1}{xx'}+\frac{1}{(1-x)(1-x')}\right)Int_{a_1a_2}(|1-n|)\nn\\
&+& \frac{k_{\perp}k_{\perp}^{'}}{xx'(1-x)(1-x')}Int_{a_1a_2}(|n|)
+\frac{|a_1-a_2|}{(x-x')^2}Int_{a_1a_2}(|1-n|)\nn\\
G_2(x,k_{\perp};x',k_{\perp}^{'})&=&
(m^2\left(\frac{1}{xx'}+\frac{1}{(1-x)(1-x')}\right)
+k_{\perp}^2\frac{1}{x(1-x)}+k_{\perp}^{'2}\frac{1}{x'(1-x')})
Int_{a_1a_2}(|n|)\nn\\
&+& k_{\perp}k_{\perp}^{'}\left(\frac{1}{xx'}Int_{a_1a_2}(|1-n|)
+\frac{1}{(1-x)(1-x')}Int_{a_1a_2}(|1+n|)\right)\nn\\
&+& \frac{|a_1-a_2|}{(x-x')^2}Int_{a_1a_2}(|n|)\nn\\
G_3(x,k_{\perp};x',k_{\perp}^{'})&=&
-\frac{m}{xx'}
\left(k_{\perp}^{'}Int_{a_1a_2}(|1+n|)
-k_{\perp}\frac{1-x'}{1-x}Int_{a_1a_2}(|n|)\right)\nn\\
G_4(x,k_{\perp};x',k_{\perp}^{'})&=&
-m^2\frac{(x-x')^2}{xx'(1-x)(1-x')}Int_{a_1a_2}(|n|)
\leea{a11}
where we have introduced the functions
\bea
&& Int_{a_1a_2}(n)=\theta(a_1-a_2)Int_{a_1}(n)
+\theta(a_2-a_1)Int_{a_2}(n)\nn\\
&& Int_{a_i}(n)=\frac{\alpha}{\pi}(-A(a_i))^{-n+1}
\left(\frac{B(a_i)}{k_{\perp}k_{\perp}^{'}}\right)^{n}\nn\\
&& A(a_i)=\frac{1}{\sqrt{a_i^2-4k_{\perp}^2k_{\perp}^{'2}}}\nn\\
&& B(a_i)=\frac{1}{2}(1-a_iA(a_i))
\leea{a12}
and the functions $a_i$, $i=1,2$ are given in \eq{a3}.

The following integrals were used by the calculation
of the matrix elements \cite{TrPa}
\bea
&& \frac{1}{2\pi}\int_0^{2\pi}d\varphi\int_0^{2\pi}d\varphi'
\frac{\cos(n(\varphi-\varphi'))}
{a_i-2k_{\perp}k'_{\perp}\cos(\varphi-\varphi')}
=2\pi(-A(a_i))^{-n+1}\left(\frac{B(a_i)}{k_{\perp}k'_{\perp}}\right)^n\nn\\
&& \frac{1}{2\pi}\int_0^{2\pi}d\varphi\int_0^{2\pi}d\varphi'
\frac{\sin(n(\varphi-\varphi'))}
{a_i-2k_{\perp}k'_{\perp}\cos(\varphi-\varphi')}=0
\leea{a13}
The last terms in $G_1,G_2$ arise from the 
$"\eta_{\mu}\eta_{\nu}"$ structure of the effective interaction \eq{a1},
and are defined only for $L_z=L'_z=0$, i.e. for 
the total angular momentum 
$J_z=S_z=\la_1/2+\la_2/2$. Due to the corresponding
$\de$-functions they are equal to
\bea
&& \De G=\frac{|a_1-a_2|}{(x-x')^2}Int_{a_1a_2}(0)
=(-\frac{\alpha}{\pi})\frac{(a_1-a_2)}{(x-x')^2}
\left(\frac{\theta(a_1-a_2)}{\sqrt{a_1^2-4k_{\perp}^2k_{\perp}^{'2}}}
-\frac{\theta(a_2-a_1)}{\sqrt{a_2^2-4k_{\perp}^2k_{\perp}^{'2}}}\right)
\leea{a14}
In the collinear limit this term is singular 
\footnote{
In fact the IR singularities in the effective interaction arise
from the singular behavior of the generator of unitary transformation
$\eta(l)$ in the collinear limit $q^+\rightarrow 0$. The following
condition must be imposed on the generator of transformation
\bea
&& lim_{q^+\rightarrow 0}~\eta(l)=0 
\leea{}
to insure the effective interaction to be finite in the collinear limit.
This is true, for example, for the second and the third choice
of the generator in Appendix C.  
}
(see also the first point of \eq{b4})
\bea
&& \De G|_{x\sim x'}=(-\frac{\alpha}{\pi})\frac{1}{x(1-x)}
\frac{1}{|x-x'|}+const
\leea{a15}
we have used
\bea
&& \hspace{-2cm}
A(a_1)\theta(a_1-a_2)-A(a_2)\theta(a_2-a_1)\nn\\
&& \hspace{-2cm} =\frac{1}{2}(A(a_1)-A(a_2))
(\theta(a_1-a_2)+\theta(a_2-a_1))
+\frac{1}{2}(A(a_1)+A(a_2))(\theta(a_1-a_2)-\theta(a_2-a_1))\nn\\
&& \hspace{-2cm} =\frac{1}{2}(A(a_1)-A(a_2))
+\frac{1}{2}(A(a_1)+A(a_2))sign(a_1-a_2)
\leea{a16}

The condition on the parameter space $(x,k_{\perp})$
due to the $\theta$-function, namely $\theta(a_1-a_2)$
(i.e. when $a_1>a_2$) reads
\bea
&& (x-x')\left(x(1-x)(k_{\perp}^{'2}+m^2)
-x'(1-x')(k_{\perp}^2+m^2)\right)>0
\leea{a17}
Making use of the coordinate change \eq{r34}
this is equivalent to
\bea
&& \left(\frac{\mu\cos\theta}{\sqrt{\mu^2+m^2}}
-\frac{\mu'\cos\theta'}{\sqrt{\mu^{'2}+m^2}}\right)
(\mu-\mu')<0
\leea{a18}

\newpage
\section{Matrix elements of the effective interaction.
Annihilation channel}
\label{B}

We repeat the same calculations for the matrix elements 
of the effective interaction in the annihilation channel. In this case
the effective interaction generated by the unitary transformation
reads \eqs{r7}{r19}
\bea
V_{LC}^{eff} &=& e^2<\ga^{\mu}\ga^{\nu}>g_{\mu\nu}^{\perp}
\left(\frac{\theta(M_0^2-M_0^{'2})}{M_0^2}
+\frac{\theta(M_0^{'2}-M_0^2)}{M_0^{'2}}\right)\nn\\
&& -e^2<\ga^{\mu}\ga^{\nu}>\eta_{\mu}\eta_{\nu}\frac{1}{P^{+2}}
\leea{}
where
\bea
&& <\ga^{\mu}\ga^{\nu}>=\frac{\bar{v}(p'_2,\la'_2)}{\sqrt{p_2^{'+}}}\ga^{\mu}
\frac{u(p'_1,\la'_1)}{\sqrt{p_1^{'+}}}
\frac{\bar{u}(p_1,\la_1)}{\sqrt{p_1^+}}\ga^{\nu}
\frac{v(p_2,\la_2)}{\sqrt{p_2^+}}P^{+2}
\leea{}
$P^+=p_1^++p_2^+$ is the total momentum; and
\bea
&& M_0^2=\frac{k_{\perp}^2+m^2}{x(1-x)}\nn\\
&& M_0^{' 2}=\frac{k_{\perp}^{' 2}+m^2}{x'(1-x')}
\leea{}
All particle momenta are depicted on \fig{1}.
Note that the energy denominators of the effective interaction
in the annihilation channel 
do not depend on the angles $\varphi,\varphi'$.

\begin{table}[htb]
\centerline{
\begin{tabular}{|c||c|}
\hline
\parbox{1.5cm}{ \[\cal M\] } & 
\parbox{12.5cm}{
\[
\frac{1}{\sqrt{k^+k^{'+}}}
\bar{v}(k',\lambda') {\cal M} u(k,\lambda)
\]
}
\\\hline\hline
$\gamma^+$ & \parbox{12.5cm}{
\[
\hspace{3cm}
2\delta^{\lambda}_{-\lambda'}
\]}
\\\hline
$\gamma^-$ &
\parbox{12.5cm}{
\[
\hspace{-1cm}
\frac{2}{k^+k^{'+}}\left[-\left(m^2-
k_{\perp}k'_{\perp}e^{+i\lambda(\varphi-\varphi')}\right)
\delta^{\lambda}_{-\lambda'}
-m\lambda\left(k'_{\perp}e^{+i\lambda \varphi'}+
k_{\perp} e^{+i\lambda\varphi}\right)
\delta^{\lambda}_{\lambda'}\right]
\]}
\\\hline
$\gamma_{\perp}^1$ & 
\parbox{12.5cm}{
\[
\left(\frac{k'_{\perp}}{k^{'+}}e^{-i\lambda\varphi'}+
\frac{k_{\perp}}{k^+}e^{+i\lambda\varphi}
\right)\delta^{\lambda}_{-\lambda'}
-m\lambda\left(\frac{1}{k^{'+}}+\frac{1}{k^+}
\right)\delta^{\lambda}_{\lambda'}
\]}
\\\hline
$\gamma_{\perp}^2$ &
\parbox{12.5cm}{
\[ 
i\lambda\left(\frac{k'_{\perp}}{k^{'+}}e^{-i\lambda\varphi'}-
\frac{k_{\perp}}{k^+}e^{+i\lambda\varphi}
\right)\delta^{\lambda}_{-\lambda'}
-im\left(\frac{1}{k^{'+}}+\frac{1}{k^+}
\right)\delta^{\lambda}_{\lambda'}
\]}\\
\hline
\end{tabular}
}
\caption[Matrix elements of the Dirac spinors]
{Matrix elements of the Dirac spinors.}
\end{table}

\subsection{The general helicity table}

For the calculation of matrix elements of effective interaction
in the annihilation channel 
we use the matrix elements of the Dirac spinors 
listed in Table $4$ \cite{LeBr}.  
Also the following holds
$ (\bar{v}_{\la'}(p)\ga^{\alpha}u_{\la}(q))^{+}
=-\bar{u}_{\la}(q)\ga^{\alpha}v_{\la'}(p)$.

We introduce
\bea
2H^{(1)}(x,\vec{k}_{\perp};\la_1,\la_2|x',\vec{k}_{\perp}^{'};\la'_1,\la'_2)
&=& <\ga^{\mu}\ga^{\nu}>g_{\mu\nu}^{\perp}
= -<\ga_1^2>-<\ga_2^2>\nn\\
2H^{(2)}(x,\vec{k}_{\perp};\la_1,\la_2|x',\vec{k}_{\perp}^{'};\la'_1,\la'_2)
&=&-<\ga^{\mu}\ga^{\nu}>\eta_{\mu}\eta_{\nu}\frac{1}{P^{+2}}
\leea{}
where
\bea
&& <\ga^{\mu}\ga^{\nu}>
=\frac{\bar{v}(1-x',-\vec{k}_{\perp}^{'};\la'_2)~\ga^{\mu}
~u(x',\vec{k}_{\perp}^{'};\la'_1)~\bar{u}(x,\vec{k}_{\perp};\la_1)~\ga^{\nu}
~v(1-x,-\vec{k}_{\perp};\la_2)}{\sqrt{xx'(1-x)(1-x')}}
\leea{}
These functions are displayed in the Table $5$.

\begin{table}[htb]
\centerline{
\begin{tabular}{|c||c|c|c|c|}\hline
\rule[-3mm]{0mm}{8mm}{\bf final:initial} & $(\lambda'_1,\lambda'_2)
=\uparrow\uparrow$ 
& $(\lambda'_1,\lambda'_2)=\uparrow\downarrow$ 
& $(\lambda'_1,\lambda'_2)=\downarrow\uparrow$ &
$(\lambda'_1,\lambda'_2)=\downarrow\downarrow$ \\ \hline\hline
\rule[-3mm]{0mm}{8mm}$(\lambda_1,\lambda_2)=\uparrow\uparrow$ & 
$H_1(1,2)$   
&$H_3(2,1)$ & $H^*_3(2,1)$ & $0$ \\ \hline
\rule[-3mm]{0mm}{8mm}$(\lambda_1,\lambda_2)=\uparrow\downarrow$ & 
$H_3(1,2)$ 
& $H^*_2(1,2)$ & $H_4(2,1)$ &$0$ \\ \hline
\rule[-3mm]{0mm}{8mm}$(\lambda_1,\lambda_2)=\downarrow\uparrow$& 
$H_3^*(1,2)$ & $H_4(1,2)$ & $H_2(1,2)$  & $0$\\ \hline
\rule[-3mm]{0mm}{8mm} $(\lambda_1,\lambda_2)=\downarrow\downarrow$ & $0$ 
& $0$ & $0$ & $0$  \\
\hline
\end{tabular}
}
\protect\caption[General helicity table defining the effective interaction
in the annihilation channel.]
{\protect\label{GeneralHelicityTableAnnihilation}General helicity table 
defining the effective interaction in the annihilation channel.}
\end{table}
\vspace{0.5cm}

Here, the matrix elements 
$H^{(n)}_i(1,2)=H^{(n)}_i(x,\vec{k}_{\perp};x',\vec{k}_{\perp}^{'})$
are the following
\bea
H^{(1)}_1(x,\vec{k}_{\perp};x',\vec{k}_{\perp}^{'})
&=&m^2\left(\frac{1}{x}+\frac{1}{1-x}\right)
\left(\frac{1}{x'}+\frac{1}{1-x'}\right)\nn\\
H^{(1)}_2(x,\vec{k}_{\perp};x',\vec{k}_{\perp}^{'})
&=&k_{\perp}k'_{\perp}\left(\frac{{\rm e}^{i(\varphi-\varphi^{'})}}{xx'}
+\frac{{\rm e}^{-i(\varphi-\varphi^{'})}}{(1-x)(1-x')}\right)\nn\\
H^{(1)}_3(x,\vec{k}_{\perp};x',\vec{k}_{\perp}^{'})
&=&m\la_1\left(\frac{1}{x'}+\frac{1}{1-x'}\right)
\frac{k_{\perp}}{1-x}{\rm e}^{i\varphi}\nn\\
H^{(1)}_4(x,\vec{k}_{\perp};x',\vec{k}_{\perp}^{'})
&=&-k_{\perp}k'_{\perp}
\left(\frac{{\rm e}^{i(\varphi-\varphi^{'})}}{x(1-x')}
+\frac{{\rm e}^{-i(\varphi-\varphi^{'})}}{x'(1-x)}\right)
\leea{}
and
\bea
&& H^{(2)}_1(x,\vec{k}_{\perp};x',\vec{k}_{\perp}^{'})
=H^{(2)}_3(x,\vec{k}_{\perp};x',\vec{k}_{\perp}^{'})=0\nn\\
&& H^{(2)}_2(x,\vec{k}_{\perp};x',\vec{k}_{\perp}^{'})
=H^{(2)}_4(x,\vec{k}_{\perp};x',\vec{k}_{\perp}^{'})=2
\leea{}

\subsection{The helicity table of the annihilation channel
for $|J_z|\leq 1$}

The matrix elements of the effective interaction
for $J_z\geq 0$~~ 
$F(x,k_{\perp};\la_1,\la_2|x',k'_{\perp};\la'_1,\la'_2)=
<x,k_{\perp};J_z,\la_1,\la_2|\tilde{V}_{LC}^{eff}|
x',k'_{\perp};J'_z,\la'_1,\la'_2>|_{ann}$ in the annihilation channel
(the sum of the generated interaction for $J_z=+1$
and instantaneous graph for $J_z=0$)
are given in Table $6$.

\begin{table}[htb]
\centerline{
\begin{tabular}{|c||c|c|c|c|}\hline
\rule[-3mm]{0mm}{8mm}{\bf final:initial} & $(\lambda'_1,\lambda'_2)
=\uparrow\uparrow$ 
& $(\lambda'_1,\lambda'_2)=\uparrow\downarrow$ 
& $(\lambda'_1,\lambda'_2)=\downarrow\uparrow$ &
$(\lambda'_1,\lambda'_2)=\downarrow\downarrow$ \\ \hline\hline
\rule[-3mm]{0mm}{8mm}$(\lambda_1,\lambda_2)=\uparrow\uparrow$ & 
$F_1(1,2)$   
&$F_3(2,1)$ & $F^*_3(2,1)$ & $0$ \\ \hline
\rule[-3mm]{0mm}{8mm}$(\lambda_1,\lambda_2)=\uparrow\downarrow$ & 
$F_3(1,2)$ 
& $F^*_2(1,2)$ & $F_4(2,1)$ &$0$ \\ \hline
\rule[-3mm]{0mm}{8mm}$(\lambda_1,\lambda_2)=\downarrow\uparrow$& 
$F_3^*(1,2)$ & $F_4(1,2)$ & $F_2(1,2)$  & $0$\\ \hline
\rule[-3mm]{0mm}{8mm} $(\lambda_1,\lambda_2)=\downarrow\downarrow$ & $0$ 
& $0$ & $0$ & $0$  \\
\hline
\end{tabular}
}
\protect\caption[Helicity table of the effective interaction
in the annihilation channel for $J_z\ge 0$]
{\protect\label{HelicityTableAnnihilation}Helicity table 
of the effective interaction
in the annihilation channel for $J_z\ge 0$.}
\end{table}
\vspace{0.5cm}

The function $F_i(1,2)=F_i(x,k_{\perp};x',k'_{\perp})$ are 
the following
\bea
F_1(x,k_{\perp};x',k_{\perp}^{'})
&=&-\frac{\alpha}{\pi}\frac{1}{\Omega}
\frac{m^2}{xx'(1-x)(1-x')}\de_{|J_z|,1}\nn\\
F_2(x,k_{\perp};x',k_{\perp}^{'})
&=&-\frac{\alpha}{\pi}\left(\frac{1}{\Omega}
\frac{k_{\perp}k'_{\perp}}{xx'}\de_{|J_z|,1}
+2\de_{J_z,0}\right)\nn\\
F_3(x,k_{\perp};x',k_{\perp}^{'})
&=&-\frac{\alpha}{\pi}\frac{1}{\Omega}
\la_1\frac{m}{x'(1-x')}
\frac{k_{\perp}}{1-x}\de_{|J_z|,1}\nn\\
F_4(x,k_{\perp};x',k_{\perp}^{'})
&=&-\frac{\alpha}{\pi}\left(-\frac{1}{\Omega}
\frac{k_{\perp}k'_{\perp}}{x(1-x')}\de_{|J_z|,1}
+2\de_{J_z,0}\right)
\leea{}
where we have introduced
\bea
&& \frac{1}{\Omega}=\frac{\theta(M_0^2-M_0^{' 2})}{M_0^2}
+\frac{\theta(M_0^{'2}-M_0^2)}{M_0^{' 2}}
\leea{}

The table for $J_z=-1$ is obtained by inverting all helicities, i.e.
\bea
&& F(J_z=+1;\la_1,\la_2)=-\la_1 F(J_z=-1;-\la_1,-\la_2)
\leea{}

The matrix elements of the effective interaction 
in the annihilation channel are nonzero only for $|J_z|\leq 1$
due to the restriction on the angular momentum of the photon.

\newpage

\section{Collinear limit}
\label{C}

We estimate in this appendix the correction of 
$"\eta_{\mu}\eta_{\nu}"$ term in the collinear limit. 
With respect to different choices of the similarity function one has 
either singular or finite corrections. 
The expansion of the effective interaction in the exchange channel
\eq{r30} with respect to
\bea
&& \frac{|\de|}{\tilde{\De}}<<1 
\leea{b1}
reads
\bea
V_{LC}^{eff} &=&
   -e^2<\ga^{\mu}\ga^{\nu}>g_{\mu\nu}\frac{1}{\tilde{\De}}
\left(1-(\Theta_1-\Theta_2)\frac{\de}{\tilde{\De}}
+(\frac{\de}{\tilde{\De}})^2
\right)\nn\\
&& -e^2<\ga^{\mu}\ga^{\nu}>\eta_{\mu}\eta_{\nu}\frac{1}{q^{+2}}
\left((\Theta_1-\Theta_2)\frac{\de}{\tilde{\De}}
+(\frac{\de}{\tilde{\De}})^2\right)
+O\left((\frac{\de}{\tilde{\De}})^3\right)
\nn\\
&=& V^{(0)}+\De V_{g_{\mu\nu}}+\De V_{\eta_{\mu}\eta_{\nu}}
\leea{b2}
where the leading order electron-positron interaction
$V^{(0)}$ is given by \eq{r35} in the main text,
the corrections from next to leading orders of
$"g_{\mu\nu}"$ and $"\eta_{\mu}\eta_{\nu}"$ terms
are called $\De V_{g_{\mu\nu}}$ and $\De V_{\eta_{\mu}\eta_{\nu}}$,
respectively; $"\Theta$-factors" are given
\bea
&& \Theta_1=\int_0^{\infty}\frac{df(\De_1)}{d\la'}f(\De_2)d\la'\nn\\
&& \Theta_2=\int_0^{\infty}\frac{df(\De_2)}{d\la'}f(\De_1)d\la'
\leea{b3}
The effective interactions with the corresponding 
similarity functions  and $"\Theta$-factors" are written below;
also the leading order corrections as $x\sim x'$
$\De V_{\eta_{\mu}\eta_{\nu}}$ are given
\bea
&& \hspace{-2cm} 
1.~~f_{\la}(\De)=\theta(\la^2-|\De|);~~
\Theta_1=\theta(\tilde{\De}_1-\tilde{\De}_2)=\theta(\de)\nn\\
&& \hspace{-2cm}
V_{LC}^{eff}=-e^2<\ga^{\mu}\ga^{\nu}>g_{\mu\nu}\left(
\frac{\theta(\tilde{\De}_1-\tilde{\De}_2)}{\tilde{\De}_1}
+\frac{\theta(\tilde{\De}_2-\tilde{\De}_1)}{\tilde{\De}_2}
\right)\nn\\
&& \hspace{-2cm}
-e^2<\ga^{\mu}\ga^{\nu}>\eta_{\mu}\eta_{\nu}\frac{1}{2q^{+2}}
|\tilde{\De}_1-\tilde{\De}_2|\left(
\frac{\theta(\tilde{\De}_1-\tilde{\De}_2)}{\tilde{\De}_1}
+\frac{\theta(\tilde{\De}_2-\tilde{\De}_1)}{\tilde{\De}_2}
\right)\nn\\
&& \hspace{-2cm}
\De V_{\eta_{\mu}\eta_{\nu}}\approx
-e^2<\ga^{\mu}\ga^{\nu}>\eta_{\mu}\eta_{\nu}\frac{1}{q^{+2}}
\frac{|\de|}{\tilde{\De}}
\approx -\frac{1}{|x-x'|}\frac{2e^2}{x(1-x)}
\frac{|k_{\perp}^2-k_{\perp}^{'2}|}
{(\vec{k}_{\perp}-\vec{k}_{\perp}^{'})^2}
+\frac{e^2}{x^2(1-x)^2}
\frac{(k_{\perp}^2-k_{\perp}^{'2})^2}
{(\vec{k}_{\perp}-\vec{k}_{\perp}^{'})^4}
\nn\\
&& \hspace{-2cm}
2.~~f_{\la}(\De)={\rm e}^{-\frac{\De^2}{\la^4}};~~
\Theta_1=\frac{\tilde{\De}_1^2}{\tilde{\De}_1^2+\tilde{\De}_2^2}\nn\\
&& \hspace{-2cm}
V_{LC}^{eff}=-e^2<\ga^{\mu}\ga^{\nu}>g_{\mu\nu}
\frac{\tilde{\De}_1+\tilde{\De}_2}{\tilde{\De}_1^2+\tilde{\De}_2^2}
-e^2<\ga^{\mu}\ga^{\nu}>\eta_{\mu}\eta_{\nu}\frac{1}{2q^{+2}}
\frac{(\tilde{\De}_1-\tilde{\De}_2)^2}{\tilde{\De}_1^2+\tilde{\De}_2^2}
\nn\\
&& \hspace{-2cm}
\De V_{\eta_{\mu}\eta_{\nu}}\approx
-e^2<\ga^{\mu}\ga^{\nu}>\eta_{\mu}\eta_{\nu}\frac{1}{q^{+2}}
\frac{\de^2}{\tilde{\De}^2}
\approx -\frac{e^2}{x^2(1-x)^2}
\frac{(k_{\perp}^2-k_{\perp}^{'2})^2}
{(\vec{k}_{\perp}-\vec{k}_{\perp}^{'})^4}\nn\\
&& \hspace{-2cm}
3.~~f_{\la}(\De)={\rm e}^{-\frac{|\De|}{\la^2}};~~
\Theta_1=\frac{\tilde{\De}_1}{\tilde{\De}_1+\tilde{\De}_2}\nn\\
&& \hspace{-2cm}
V_{LC}^{eff}=V^{PT}_{\la=\La\rightarrow\infty}+V^{inst}
=-e^2<\ga^{\mu}\ga^{\nu}>g_{\mu\nu}\frac{1}{\tilde{\De}}\nn\\
&& \hspace{-2cm}
\De V_{\eta_{\mu}\eta_{\nu}}=0
\leea{b4}

\newpage

\end{document}